\documentclass{aastex631}
\usepackage{float}
\usepackage{gensymb}
\usepackage{graphicx}
\usepackage{amsmath}
\usepackage{tabularx}
\usepackage{multirow}
\usepackage{esvect}
\usepackage{rotating}

\received{August 8, 2023}
\shorttitle{\emph{NuSTAR} observations of Pictor A}
\shortauthors{Shaik et al.}
\revised{May 6, 2024}
\accepted{May 21, 2024}

\begin{document}

\title{Characterization of the western Pictor A hotspot in the hard X-rays with \emph{NuSTAR} }

\author{Aamil Shaik}
\affiliation{Department of Physics\\
University of Maryland Baltimore County\\
1000 Hilltop Circle Baltimore\\
MD 21250, USA}

\author[0000-0002-7676-9962]{Eileen T. Meyer}
\affiliation{Department of Physics\\
University of Maryland Baltimore County\\
1000 Hilltop Circle Baltimore\\
MD 21250, USA}

\author[0000-0001-9018-9553]{Karthik Reddy}
\affiliation{Department of Physics\\
University of Maryland Baltimore County\\
1000 Hilltop Circle Baltimore\\
MD 21250, USA}

\author[0000-0003-2714-0487]{Sibasish Laha} 

\affiliation{Center for Space Science and Technology, University of Maryland Baltimore County, 1000 Hilltop Circle, Baltimore, MD 21250, USA.}
\affiliation{Astrophysics Science Division, NASA Goddard Space Flight Center, Greenbelt, MD 20771, USA.}
\affiliation{Center for Research and Exploration in Space Science and Technology, NASA/GSFC, Greenbelt, Maryland 20771, USA}

\author[0000-0002-2040-8666]{Markos Georganopoulos}
\affiliation{Department of Physics\\
University of Maryland Baltimore County\\
1000 Hilltop Circle Baltimore\\
MD 21250, USA}

\begin{abstract}
  The origin of X-ray emission from the resolved kiloparsec-scale jets and hotspots of many active galactic nuclei (AGN) remains uncertain, particularly where the X-ray emission is separate from the radio-optical synchrotron component. Possible explanations include synchrotron emission from a second electron population and external Compton or synchrotron self-Compton processes -- alternatives which imply very different physical conditions within the jet. Until recently, X-ray studies of resolved jets and hotspots have been restricted to below $\sim$10 keV, often showing a hard spectral index indicating a spectral peak beyond this energy range. Here we present NuSTAR observations of the nearby powerful radio galaxy Pictor A, in which we clearly detect the western hotspot at approximately 4' from the host galaxy, the most significant detection of hotspot emission above 10 keV to date. The NuSTAR spectrum is best fit by a single powerlaw of index $\Gamma = 2.03\pm0.04$; an exponential cut-off gives a 1$\sigma$ lower limit on the cutoff energy of 40.7 keV. We confirm previous findings of variations in the soft X-ray flux detected by \emph{Chandra} over the 2000 to 2015 period, at a significance of 6.5$\sigma$. This rises to $>8\sigma$ in the common 3-8 keV band using the combined 22-year span of \emph{Chandra} and \emph{NuSTAR} observations. The variability of the western Pictor A hotspot strongly confirms the previously argued synchrotron nature of the X-ray emission for the hotspot, while the lower bound to the spectral cutoff energy implies electron energies in the hotspot reach up to at least a few TeV.
\end{abstract}

\keywords{X-rays: galaxies -- galaxies: active -- galaxies: jets -- quasars: individual (Pictor A)}

\section{Introduction} \label{sec:intro}

A fraction of active galactic nuclei (AGN) launch large, collimated
jets of relativistic plasma \citep{BLANDFORD1979}. These jets can
transport energy and matter up to Megaparsec distances from the
central black hole in the most extreme cases \citep{dab2020}. In more
powerful jets, the relativistic jet can extend into the intergalactic
medium (IGM), creating a bright, terminal hotspot where the kinetic
energy of the jet is injected into large, extended lobes of cooling
plasma. The power and extent of AGN jets is such that they can produce
Mpc-scale shocks and cavities in the IGM and the intercluster medium
and are thought to have a significant impact on the star-formation
rate of the host galaxy \citep{MCNAMARA2009A,FABI2012}.

AGN jets and hotspots are both primarily detected at radio and
sometimes IR/optical frequencies where they emit synchrotron radiation
from electrons gyrating within the jet magnetic field \citep[e.g.,
][]{MEISENHEIMER1997,WERNER2012}. For this reason, jetted AGN are more
`radio-loud' compared to the majority of non-jetted AGN. Sources where
the jet is aligned in the plane of the sky are known as radio
galaxies. Radio galaxies are often divided into two classifications
based on jet morphology: Fanaroff $\&$ Riley type I \citep[FR
  I;][]{FANA1974} jets tend to be shorter, less collimated, and
brightest near the central engine with the jet typically terminating
in large, darkened plumes of plasma, while FRII jets are highly
collimated and brightest and exhibit terminal hotspots.

Low-spatial-resolution photometric observations of jetted AGN are
usually dominated by emission from the highly relativistic base of the
jet (sometimes called the `core' due to appearance in e.g., radio-band
imaging). The focus of this study is on the emission of the extended
jet, from $\sim0.1-100$ kpc from the central engine. At these scales,
the kpc-scale jet extends into and beyond the host galaxy and can be
resolved in high-resolution imaging with e.g. the Hubble Space
Telescope, the Very Large Array, and the \emph{Chandra} X-ray
Observatory.

One of the major discoveries of \emph{Chandra} was the detection of
bright X-ray emission from resolved radio jets and hotspots on kpc
scales \citep{CHARTAS2000,SAMBRUNA2001}. In some FRI jets lacking
hotspots such as M87 \citep{HARRIS2003,SUN2018} and Cen A
\citep{KRAFT2002,SNIOS2019}, the X-ray emission is roughly consistent
with a single spectral component extending from radio to optical to
X-rays. In other, often more powerful jets
\citep{HARRIS2006,HARRIS2007,HARRIS2010,MARSHALL2010} the X-ray
emission is often `anomalous', meaning that the X-ray flux from the
jet and terminal hotspot is too bright and hard-spectrum to be an
extension of the primary radio-optical synchrotron component.

For kpc-scale anomalous jets (here we use 'jet' to mean the resolved
components extending up to but not including the hotspot), the
initially preferred explanation of inverse Compton scattering of the
CMB (IC-CMB) has largely been overturned in favor of a synchrotron
origin for the X-rays, at least in jets a lower redshift
\citep[e.g.][]{2000TAVECCHIO,2001CELOTTI,SAMBRUNA2004,JESTER2006,
  Meyer2016, BREI2017, BREIDING2023, MEYER2023}.

In hotspots the case is somewhat different; thought to represent the
jet terminus where the flow finally decelerates to sub-relativistic
bulk speed, the IC-CMB contributions are expected to be
negligible. While many hotspots are adequately explained as
synchrotron self-Compton (SSC) emission \citep{harris1994,
  hardcastle2002, ARSHAKIAN2000}, others are not, requiring magnetic
fields grossly out of equipartition \citep{hardcastle2004}. An SSC
origin is also incompatible with the sometimes observed large spatial
offsets between the radio and X-rays \citep{ERLUND2007,Reddy_2023}.
Beaming models based on jet electrons upstream of the termination
shock where the flow is still relativistic scattering post-shock radio
emission have been proposed \citep{GEOR2003,WORRALL2012}, but there is
limited evidence to support the conclusion that beaming strongly
impacts hotspot X-ray emission across the jet population
\citep[e.g.][]{MULLIN2008,HARDCASTLE2016} and more recent work
suggests that the X-ray emission in many hotspots is due synchrotron
emission from a second population of electrons, as has been found for
many jets \citep[e.g.][]{MINGO2017}.

The source of the kpc-scale X-ray emission for powerful jets and
hotspots is a longstanding open question. In some cases alternate
models imply orders-of-magnitude differences in jet kinetic power and
magnetic field strength. Determining to what extent each possible
mechanism contributes across the X-ray jet and hotspot population is
thus important also for understanding the energy content of jets and
their impact. Given the unsolved origin of the X-ray emission, it is
useful to attempt to extend the spectral coverage with observations in
the hard X-ray range ($E > 10$ keV). If the emission is synchrotron in
origin, locating the turnover point where the spectrum begins to fall
can place constraints on the maximum electron energy. This energy
range is covered by \emph{NuSTAR}, but observations of kpc-scale jets
are complicated by the fact that \emph{NuSTAR} lacks the angular
resolution (FWHM$\sim$7.5") to easily distinguish the kpc-scale jet
from the core in most sources. Given these limitations, the outer jet
and western hotspot of FR II radio galaxy Pictor A ($z = 0.035$; 0.69
kpc/") is nearly the only jet with the sufficient angular separation
($\sim4'$) from the core to be a viable target for hard X-ray study
and the only one bright enough to be detected by \emph{NuSTAR} in less
than approximately 1 Ms of exposure.

Pictor A is among the most well-studied X-ray jets in the literature,
and, along with its hotspot, has been the subject of a number of
detailed studies over the past two decades
\citep[e.g.,][]{WILSON2001,hardcastle2004,HARDCASTLE2005,MIGLIORI2007,TINGAY2008,MARSHALL2010,HARDCASTLE2016,THIMMAPPA2020}. The
X-ray jet was first reported by \cite{WILSON2001} and was noted to be
X-ray-anomalous by \cite{hardcastle2004}. Despite its bright X-ray
emission, it is not a source which we would expect a significant
IC-CMB component as it has a low redshift, its (approaching) western
jet is not closely inclined to our line-of-sight
($\theta\sim20-45\degree$), and its counterjet is clearly visible
\citep{HARDCASTLE2005}. A beamed inverse-Compton origin for the X-rays
from the jet or hotspot would require magnetic field strengths
significantly below equipartition values \citep{hardcastle2004}. The
deepest observation of the jet and hotspot to date in the soft X-rays
combined 464 ks of \emph{Chandra} exposure on the western hotspot,
finding a spectrum near-equally consistent with a single constant
powerlaw spectrum with $\Gamma = 1.94\pm0.01$ or a steepening, broken
powerlaw spectrum with $\Gamma_{1} = 1.86\pm0.02$, $\Gamma_{2} =
2.16\pm0.05$, and a break energy at around 2 keV
\citep{HARDCASTLE2016}. One of the motivations for our deep NuSTAR
observation was to clarify the presence and location of a possible
spectral break or cutoff in the broad X-ray
band. Previous research of the hard X-ray emission
  from the hotspot found no evidence of such a spectral break or
  cutoff \citep{SUNADA2022}; however, as discussed below, the addition
  of new \emph{NuSTAR} observations affords us the opportunity to
  reexamine the spectrum with greater significance.

The jet of Pic A is also noteworthy for the detection of variability
in the X-rays. In 2010, \cite{MARSHALL2010} analyzed three
\emph{Chandra} observations taken in 2000, 2002, and 2009 of 26, 96,
and 54 ks of exposure time respectively. They found that the soft
X-ray flux of two features located at 34 and 49 kpc away from the core
decreased significantly ($3.4\sigma$ and $\sim3\sigma$ respectively)
over the course of the observations. The later \cite{HARDCASTLE2016}
study (with an additional 6 \emph{Chandra} observations) found the
hotspot X-ray flux decreased by $\sim10\%$ over a time period of
approximately one month, and confirmed $\sim$3$\sigma$ variability in
the jet. Recently, Pictor A was included in a large archival study of
\emph{Chandra} jet observations where at least five regions in the
jet, including the hotspot, were found to be variable
\citep{MEYER2023}; we further expand on the results of that study here
using the same methods. The IC-CMB mechanism requires low-energy
electrons ($\gamma\sim100-1000$) with cooling timescales on the order
of $\sim10^{5}-10^{6}$ yrs \citep{WORRALL2009}, so the observed
variability in the X-ray flux is clearly incompatible with such an
origin. Furthermore, the fast timescales of the variability allow us
to probe the physical conditions within the hotspot, most notably the
size of the emitting regions \citep{TINGAY2008}.

In this paper, we report the results of a comprehensive X-ray spectral
analysis of the western hotspot of Pictor A, including new and
archival \emph{NuSTAR} observations totalling 348 ks. We combine this
with 313 ks of archival \emph{Chandra} data from 11 different
observations with the Advanced CCD Imaging Spectrometer
\citep[ACIS;][]{GARMIRE2003} and modeled the combined wide X-ray band
spectrum ($0.5 - 20$ keV). A similar research
  project was undertaken by \cite{SUNADA2022}; however, access to an
  additional 300 ks of \emph{NuSTAR} data allows us to place stricter
  constraints on the hard and broadband X-ray spectrum than previously
  possible. We also used the maximum likelihood method of
\cite{MEYER2023} to test for variability in the \emph{NuSTAR} and
\emph{Chandra} observations, and discuss the implications of our
findings for the X-ray emission origin in Pictor A.

\begin{figure*}[!th]
    %\centeringl
    \gridline{
     \fig{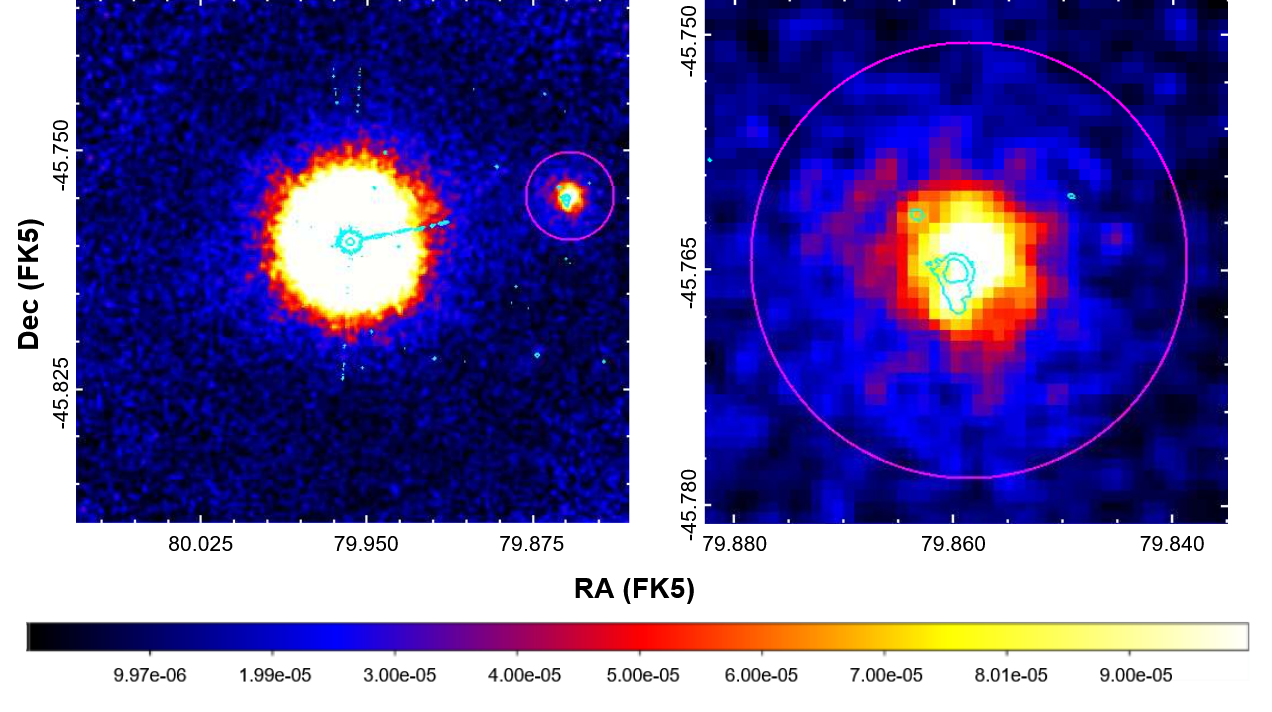}{0.85\textwidth}{hbt!}
    }
    \caption{(Left) The final combined \emph{NuSTAR} image of Pictor A using FPMA and FPMB data from epochs 1 and 2. The images from each observation are aligned using cross-correlations of the bright core. The images are then background-subtracted and exposure-corrected using background- and exposure-maps generated by \texttt{nuskybgd} and \texttt{nuproducts}, respectively. The individual flux images are then averaged, and smoothed by a Gaussian at a factor of 2 pixels (4.8") to produce the image above. The 50" extraction region used for spectral analysis is encircled in magenta. The overlaid contours are taken from combined \emph{Chandra} data. (Right) An expanded image of the source region and western hotspot, again with \emph{Chandra} contours overlaid. Unlike the high-resolution imaging of \emph{Chandra}, \emph{NuSTAR} is unable to resolve any structures or morphological features within the hotspot. The colorbar below shows the image flux scale in units of ph s$^{-1}$ cm$^{-2}$.}
    \label{fig:nu764X}
\end{figure*}

\section{Observations and data reduction}

\subsection{\emph{NuSTAR} observations}

The \emph{NuSTAR} observatory consists of two independent, co-aligned
X-ray telescopes, and was launched into orbit in 2012
\citep{HARRISON2013,MADSEN2014}. The telescopes each have an array of
grazing incidence X-ray mirrors arranged in a conical approximation of
a Wolter I geometry, which are designated as Optical Module A (OMA)
and Optical Module B (OMB). OMA and OMB focus incoming X-rays within
the $13'\times13'$ FOV onto two modules of solid state detector
arrays, which are designated as Focal Plane Module A (FPMA) and Focal
Plane Module B (FPMB).

\emph{NuSTAR} first observed the western hotspot of Pictor A in 2015
with an exposure of 109 ks. A second observation was proposed and
completed in 2021, consisting of 3 pointings totaling 302 ks of
exposure. We group these later observations into a single `epoch', or
collection of observations conducted within a short timeframe of one
another ($t_{sep} \leq$ 1 wk). The details of the 2 \emph{NuSTAR}
epochs and their constituent observations which are used in this study
are included in Table \ref{tab:obs}, which lists the project code,
observation date, PI, the epoch grouping, and the exposure time before
and after filtering.

\begin{table}[h!]
  \caption{\emph{NuSTAR} and \emph{Chandra} observations of Pictor A }
\centering
 \begin{tabular}{l|cccccl}
   \hline
   Instrument & Epoch & Project & Date Obs. & Exposure & Filtered & PI \\ 
&  & Code & (YYYY-MM-DD) & (ks) & Exposure (ks) &   \\
   \hline\hline
    \emph{NuSTAR} & 1 & 60101047002 & 2015-12-03 & 109.5 & 92.82 & Croston \\
        & 2 & 60701064002 & 2021-11-28 & 126.2 & 104.1 & Meyer \\
        & 2 & 60701064004 & 2021-12-01 & 37.59 & 32.39 & Meyer \\
        & 2 & 60701064006 & 2021-12-02 & 138.8 & 118.6 & Meyer \\
        & Total & - & - & 412.1 & 347.9 & - \\
        &&&&&& \\
        \emph{Chandra} & 1 & 346 & 2000-01-18 & 25.83 & 25.07 & Wilson \\
        & 2 & 3090 & 2002-09-17 & 46.36 & 31.23 & Wilson \\
        & 2 & 4369 & 2002-09-22 & 49.12 & 34.53 & Wilson \\
        & 3 & 11586 & 2009-12-12 & 14.26 & 12.21 & Hardcastle \\
        & 3 & 12039 & 2009-12-07 & 23.74 & 21.95 & Hardcastle \\
        & 3 & 12040 & 2009-12-09 & 17.32 & 15.62 & Hardcastle \\
        & 4 & 14221 & 2012-11-06 & 37.48 & 35.18 & Hardcastle \\
        & 4 & 15580 & 2012-11-08 & 10.48 & 9.97 & Hardcastle \\
        & 5 & 14222 & 2014-01-17 & 45.38 & 44.03 & Hardcastle \\
        & 6 & 14223 & 2014-04-21 & 50.14 & 42.20 & Hardcastle \\
        & 7 & 16478 & 2015-01-09 & 26.82 & 23.75 & Hardcastle \\
        & 7 & 17574 & 2015-01-10 & 18.62 & 17.34 & Hardcastle \\
        & Total & - & -          & 365.6 & 313.1 & - \\
 \hline
 \end{tabular}
 \label{tab:obs}
\end{table}

We downloaded the raw \emph{NuSTAR} data from \emph{HEASARC} and
applied the standardized procedure to reduce and image the data using
the \texttt{nupipeline} command from the \emph{NuSTAR} Data Analysis
Software (\emph{NuSTARDAS})
v2.1.0\footnote{https://heasarc.gsfc.nasa.gov/docs/nustar/analysis/nustar\_swguide.pdf},
which is included in the \emph{HEASARC} software package
(\emph{HEAsoft})
v6.29\footnote{https://heasarc.gsfc.nasa.gov/docs/software/heasoft/developers\_guide/}. In
particular, we excised time intervals where background radiation and
contamination from the South Atlantic Anomaly (SAA) was significant
using the \texttt{nupipeline} settings \texttt{SAA = STRICT} and
\texttt{TENTACLE = YES}. At this stage, we filtered the data to the
energy range of 3-78 keV.

Once the data were filtered, we created a circular extraction region
of radius 50" at the centroid of the hotspot. We used the
\texttt{nuproducts} pipeline to generate high-level data products for
the source region, including extracted photon counts spectral files
(\texttt{PHA}s), auxiliary response files (\texttt{ARF}s), response
matrix files (\texttt{RMF}s), and exposure maps. This process was
repeated for both the FPMA and FPMB images from each \emph{NuSTAR}
observation, giving us 8 total sets of \emph{NuSTAR} data.

For visual study of the jet and environs, we created a hard X-ray
image using the combined data of all available \emph{NuSTAR}
observations. We used the \texttt{nuskybgd} code developed by
\cite{WIK2014} to generate background-subtracted images (see Section
\ref{sec:nubgd} for more details). We used the background-subtracted
images and the \texttt{CIAO} commands \texttt{dmstat} and
\texttt{acrosscorr} to calculate the centroid position of the
auto-correlated image of the earliest FPMA observation of the X-ray
core, calculating the offsets in the centroid positions of later
observations, and reprojecting the later observations such that the
X-ray core is aligned. We used the same offsets to align the exposure
maps generated by \texttt{nuproducts}. We co-added all our aligned
\emph{NuSTAR} observations and divided the resulting deep photon image
by the similarly co-added exposure map to obtain a hard X-ray flux
image of Pictor A. Lastly, we applied a Gaussian smoothing filter on
the image at a factor of 2 pixels (4.8") to produce our final image,
which is shown in Figure \ref{fig:nu764X}, with \emph{Chandra}
contours overlaid.

\textcolor{black}{The peak of the hotspot emission profile as seen by
  \emph{NuSTAR} at first appears slightly offset from the peak
  observed by \emph{Chandra}, by approximately 4-5$''$ downstream
  along the jet axis. If real, this offset would place the hard X-ray
  centroid beyond even the radio peak, which is estimated to be
  downstream of the \emph{Chandra} position by approximately 1$''$
  \citep{HARDCASTLE2016}. However, there are two factors which lead us
  to doubt the reality of the offset: first, there is substantial
  uncertainty in the centroid position, with dx, dy = 1.72$''$,
  2.42$''$ respectively, where we use the root mean square correction
  obtained from cross-correlations on the \textit{NuSTAR} images with
  the \texttt{acrosscorr} command. In addition, asymmetries in the
  effective PSF may bias the apparent position of the hotspot --
  indeed deconvolved \emph{NuSTAR} images do not show an offset (see
  further discusion and images in section\ref{deconv}). }

\subsection{\emph{Chandra} observations}

The \emph{Chandra} X-ray Observatory was launched in 1999 as a part of
NASA’s Great Observatories program \citep{WEISSKOPF2000A}. The ACIS-S
CCD detector \citep{GARMIRE2003} onboard has an unrivaled subarcsecond
angular resolution in the 0.3 - 10.0 keV energy range, making it ideal
for observing kiloparsec-scale extragalactic X-ray
jets. \emph{Chandra} has observed Pictor A in 16 separate pointings,
from the initial discovery of the large-scale X-ray jet in 2000
\citep{WILSON2001} to subsequent observations of the jet and hotspot
as recently as 2015
\citep{hardcastle2004,HARDCASTLE2005,HARDCASTLE2016}. A few
observations were excluded from this analysis due to extremely short
exposure. Similar to our \emph{NuSTAR} observations, we grouped
observations conducted within $\sim1$ week of one another into single
epochs. Ultimately, we made use of use 11 \emph{Chandra} observations
which are grouped into 7 epochs and totaling a nominal 360 ks of
exposure time. A full list of the \emph{Chandra} observations used in
this analysis, their epoch groupings, PI, project code, observation
date, and exposure time before and after filtering can be found in
Table \ref{tab:obs}, alongside the \emph{NuSTAR} epochs and
observations described in the previous section.

Two observations included in previous deep surveys of Pictor A
\citep{HARDCASTLE2016,SUNADA2022} have been excluded from this
study. The observation corresponding to project code 14357 has the
hotspot positioned on a separate chip from the rest of the
observation. To minimize complications of differences between the
chips, it has been excluded. Similarly, in the observation
corresponding to project code 15593, the hotspot is positioned exactly
on the chip edge. This complicates the extraction of spectra from the
hotspot, and so this observation was also excluded from out analysis.

All observations were reduced using \texttt{CIAO} version 4.1.2
\citep{FRUSCIONE2006} and the calibration database \texttt{CALDB}
v. 4.9.2. Each observation was first reprocessed using the
\texttt{chandra\_repro} and filtered to remove all photons outside of
the 0.4-8 keV range to minimize background contamination. We use the
\texttt{lc\_clean} task to bin the good time interval (GTI) data with
a bin time of 259.28 s and excise intervals with background flares,
which we define as a photon count rate which deviates from the mean
count rate by more than 2$\sigma$. We then used this filtered data to
produce the 11 sub-pixel images centered on the core of the X-ray jet,
one for each epoch, downsizing by a factor of 5 for a pixel scale of
0.0984". After filtering, the total exposure time is approximately 313
ks.

\begin{figure*}[!th]
    \gridline{
    \fig{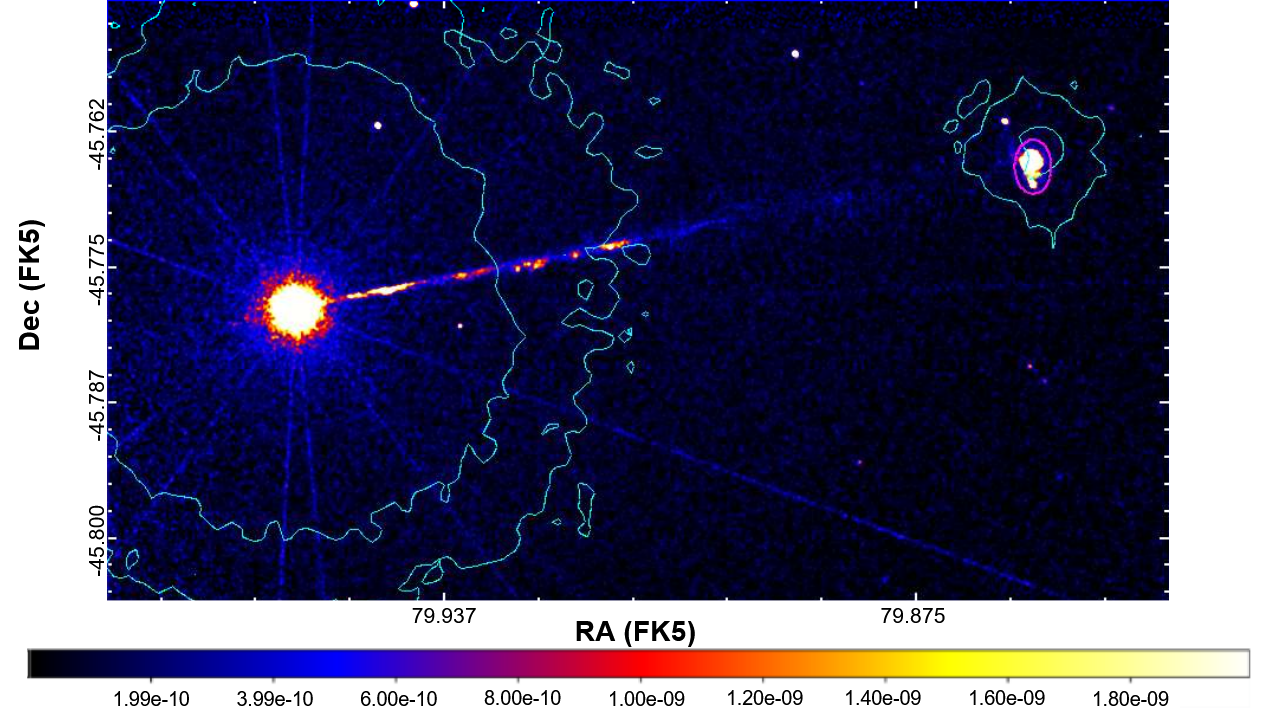}{0.85\textwidth}{}
    }
    \caption{ The final combined \emph{Chandra} image of Pictor A
      using soft X-ray data from all epochs listed in Table
      \ref{tab:obs}. The images from each observation are aligned
      using PSF simulations and cross-correlation of the bright
      core. The images are then exposure-corrected using exposure-maps
      generated by the \texttt{CIAO} tool \texttt{mkexpmap}. The
      individual flux images are then averaged together, and the final
      image is smoothed by a Gaussian at a factor of 8 pixels
      (3.92"). The elliptical extraction region used for spectral
      analysis is encircled in magenta. The overlaid cyan contours are
      taken from the combined \emph{NuSTAR} data shown in Figure
      \ref{fig:nu764X}. The colorbar below shows the image flux scale
      in units of ph s$^{-1}$ cm$^{-2}$.}
    \label{fig:chanm}
\end{figure*}

We also generated a combined soft X-ray image using all \emph{Chandra}
data reduced for this study. We again used cross-correlation to
precisely locate the hotspot centroid and structure to correct for
astrometric offsets between observations and the exposure maps
generated for each observation using the \texttt{CIAO} tool
\texttt{mkexpmap}. Finally, we combined the aligned images for each
\emph{Chandra} epoch using the \texttt{CIAO} tool \texttt{dmmerge} and
divided by the combined exposure map to obtain the final soft X-ray
flux image of Pictor A. The final image is presented in Figure
\ref{fig:chanm}, with \emph{NuSTAR} contours overlaid.

\subsection{\emph{NuSTAR} background} \label{sec:nubgd}
The background radiation spectrum for \emph{NuSTAR} observations is
complex, including multiple different components which dominate at
different energies. Additionally, the background radiation is
spatially nonuniform and varies significantly according to the
position on the detector.

\begin{figure*}[th!]
    %\centeringl
    \gridline{
    \fig{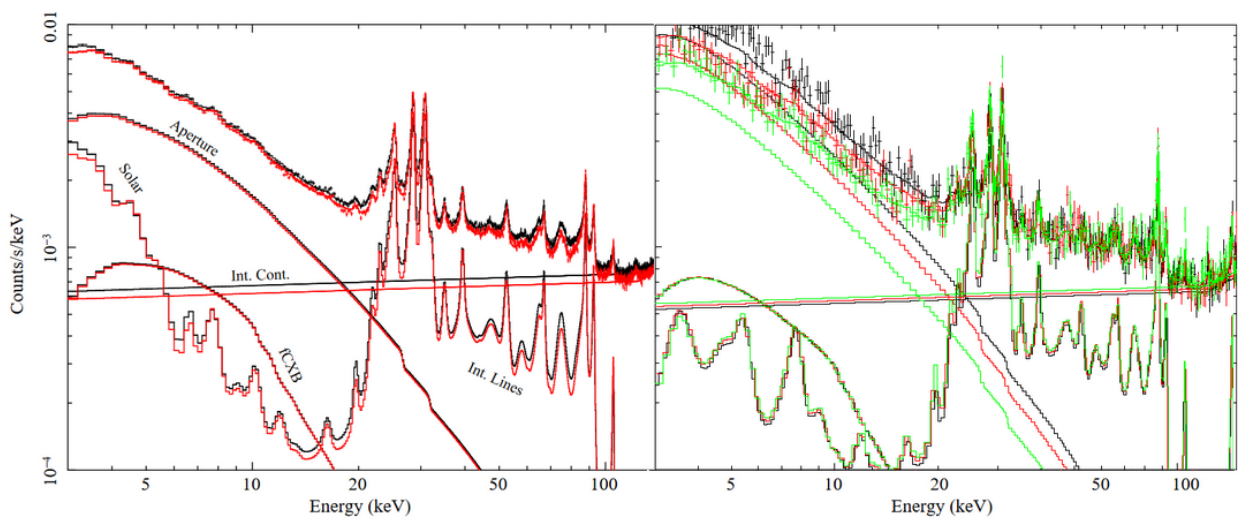}{0.80\textwidth}{}
    }
    \caption{(Left) Image of the various background components and
      total background spectrum for \emph{NuSTAR}, taken from
      \cite{WIK2014}. The model consists of an aperture background due
      to stray light leaking into the detector, an internal background
      consisting of a broken powerlaw and a spectral line continuum, a
      background of scattered light, and a background due to
      unresolved X-ray sources in the FOV. (Right) Image of the
      background fit completed using \texttt{nuskybgd}. The code fits
      background spectra from multiple annular regions which
      effectively encompass the entirety of the image which contains
      negligible source emission. The spectra from these different
      annuli are depicted as the red, green, and black spectra, which
      are binned to a minimum of 30 counts per bin, fit
      simultaneously, and used to constrain the spectral parameters
      and map the spatial variations of the background across the
      detector.}
    \label{fig:nubgmod}
\end{figure*}

The background spectra for each module and pointing were modeled and
fit using the IDL-based \texttt{nuskybgd} code developed by
\cite{WIK2014}, which maps both the spectral and spatial variations of
the background emission across the entire image. The full background
model is detailed in Appendix \ref{app:nubgd} and the best-fit
normalization for each variable parameter for each \emph{NuSTAR}
observation is recorded in Supplemental Table 1 in the online version
of this article. For all observations, $\chi^{2}_{red} = \chi^{2}$/DOF
had a value between 0.9 and 1.1, indicating that the background was
well fit by the model.  Figure \ref{fig:nubgmod} shows a visual
representation of the complete background model, including the
different individual components, taken from \cite{WIK2014} alongside
the background spectra observed in one of our \emph{NuSTAR}
observations.

We used \texttt{nuskybgd} to extract and fit the background spectrum
and make a map of the background emission over the detector
coordinates for each \emph{NuSTAR} observation. After fitting the
background, we provided \texttt{nuskybgd} with our source region and
created a model file containing the appropriate parameter values for
the background spectrum at those detector coordinates.

\subsection{\emph{Chandra} background} \label{sec:chanbgd}
For each \emph{Chandra} observation, we extracted the background
photon counts from a large elliptical region far from the hotspot and
bright X-ray core with minor and major axes of 70" and 65"
respectively. The background region position was identical across all
\emph{Chandra} epochs and always on the same chip as the hotspot. We
used this to generate a background spectral file corresponding to each
\emph{Chandra} observation of the hotspot. Whenever fitting
\emph{Chandra} data in \texttt{XSPEC}, we use the \texttt{CSTAT}
statistic which defaults to \texttt{WSTAT} when background data is
loaded but not modeled in XSPEC. We bin the spectra such that there
are at least 5 counts in the background spectra for each bin to avoid
any problematic biases.

\subsection{Spectral modeling of the western hotspot}
We used \texttt{XSPEC} v12.10.1f and its python interface PyXspec v
2.1.1 for all X-ray spectral analysis. We chose the Poisson-based
\texttt{cstat} statistic \citep{CASH1979} for our parameter estimation
statistic and the Anderson-Darling (ad) test statistic to evaluate
goodness-of-fit. We chose this method to avoid the problematic biases
that arise from trying to bin and fit the spectrum using the
$\chi^{2}$ statistic \citep{ANDRAE2010}. We fit the spectra from
\emph{Chandra} and \emph{NuSTAR} separately, then performed a final
fit using the combined data set.

For \emph{NuSTAR} observations, we use a 50" radius circular region
with its center located at the centroid of the western hotspot as our
source region. We extracted the source counts from this region for
both the FPMA and FPMB image and filtered to an energy range of 3-20
keV. Based on the results of our analysis using \texttt{nuskybgd}
across the 2 epochs of \emph{NuSTAR} data, we estimate that
approximately 3192 counts out of 3542 total counts ($\sim90\%$) in the
energy band of 20-78 keV are the result of background
radiation. Therefore, we ignored this higher energy band in our
analysis.

When modeling the source spectrum for each \emph{NuSTAR} epoch, we
combined FPMA and FPMB data, and introduced a constant factor to our
spectral model (\texttt{constant} parameter). For the FPMA
observations, this constant was fixed to 1, while for FPMB
observations, it was free to vary between 0.9 and 1.1. This factor was
included to account for the unknown normalization difference between
the two detectors which typically varies between $\pm5\%$
\citep{MADSEN2015}. After fitting each \emph{NuSTAR} epoch separately,
we performed a total fit using all reduced \emph{NuSTAR} data.

For \emph{Chandra} observations, we extracted the source counts from
an elliptical region with minor and major axes of 6" and 9"
respectively and its center located at the centroid of the western
hotspot. Just as we did with \emph{NuSTAR}, we fit each \emph{Chandra}
epoch separately, and then fit the combined \emph{Chandra} data in an
energy range of 0.5-7 keV.
\begin{table}
  \caption{Model Definitions and Bayesian Prior Distributions}
\centering
 \begin{tabular}{lll|ccccc}
Model & \texttt{XSPEC} & Model Formula & Parameter & Param. & Prior Type & Prior & Prior \\ 
Name & Model & [$F(E)$] & Name  & No. &  & Min. & Max. \\
\hline
\hline
        \textcolor{black}{Powerlaw (PL)*}  & \texttt{powerlaw} & $NE^{-\Gamma}$ & \textcolor{black}{N ($N_{1-kev}$)} & 2 & Log Unif. & -10 & 0 \\
        & & & \textcolor{black}{PhoIndex ($\Gamma$)} & 1 & Unif. & 1 & 3 \\
        & & & & & & & \\
        \multirow{2}{2cm}{\textcolor{black}{Broken powerlaw (Bkn. PL)}} & \texttt{bknpower} & $NE^{-\Gamma_{1}}, E < E_{brk}$ & N & 4 & Log Unif. & -10 & 0 \\
        & & $NE_{brk}^{\Gamma_{2}-\Gamma_{1}}E^{-\Gamma_{2}}, E > E_{brk}$ & \textcolor{black}{PhoIndx1 ($\Gamma_{1}$)} & 1 & Unif. & 1 & 3 \\
        & & & \textcolor{black}{BreakE ($E_{brk}$)} & 2 & Unif. & $E_{min}$ & $E_{max}$ \\
        & & & \textcolor{black}{PhoIndx2 ($\Gamma_{2}$)} & 3 & Unif. & 1 & 3 \\
        & & & & & & & \\
        \multirow{3}{2cm}{High-energy cutoff powerlaw} & \texttt{cutoffpl} & $NE^{-\Gamma}\mathrm{exp}(-E/\beta)$ & N & 3 & Log Unif. & -10 & 0 \\
        & & & PhoIndex & 1 & Unif. & 1 & 3 \\
        & & & HighECutoff & 2 & Unif. & $E_{min}^{*}$ & $E_{max}$\\
        & & & & & & & \\
        Log-parabola & \texttt{logpar} & $NE^{-\alpha-\beta\mathrm{log}(E)}$ & N & 4 & Log Unif. & -10 & 0 \\
        & & & alpha & 1 & Unif. & 1 & 3 \\
        & & & beta & 2 & Unif. & 0 & 1 \\
        & & & & & & & \\
        \multirow{2}{2cm}{Thermal plasma} & \texttt{apec} & \cite{SMITH2001} & N & 16 & Log Unif. & -10 & 0 \\
        & & & kT & 1 & Unif. & 0 & 30 \\
        & & & & & & & \\
        \multirow{3}{2cm}{\textcolor{black}{Double broken powerlaw (Dbl. Bkn. PL)}} & \texttt{bkn2power} & $NE^{-\Gamma_{1}}, E < E_{brk,1}$ & N & 6 & Log Unif. & -10 & 0 \\
        & & $NE_{brk,1}^{\Gamma_{2}-\Gamma_{1}}E^{-\Gamma_{2}}, E_{brk,1} < E < E_{brk,2}$ & PhoIndx1 & 1 & Unif. & 1 & 3 \\
        & & $NE_{brk,1}^{\Gamma_{2}-\Gamma_{1}}E_{brk,2}^{\Gamma_{3}-\Gamma_{2}}E^{-\Gamma_{3}}, E > E_{brk,2}$ & \textcolor{black}{BreakE1 ($E_{brk,1}$)} & 2 & Unif. & $E_{min}$ & 5 \\
        & & & \textcolor{black}{PhoIndx2 ($\Gamma_{2}$)} & 3 & Unif. & 0 & 4  \\
        & & & \textcolor{black}{BreakE2 ($E_{brk,2}$)} & 4 & Unif. & 5 & $E_{max}$ \\        
        & & & \textcolor{black}{PhoIndx3 ($\Gamma_{3}$)} & 5 & Unif. & 1 & 3 \\
        
   \hline
   \end{tabular}
         \label{tab:priors}
\tablecomments{   $^E_{min}-E_{max}$ is 0.5-8 keV for \emph{Chandra}, 3-20 keV for \emph{NuSTAR}}
\tablecomments{   \textcolor{black}{* Abbreviations are provided for the most noteworthy models and their respective parameters which are used to reference them in future tables.} }
\end{table}

Finally, to maximize our photon statistics and achieve the best
constraints on the X-ray spectrum over the wide X-ray band (0.5-20
keV), we fit the combined spectrum of \emph{Chandra} and \emph{NuSTAR}
data. We reintroduce the constant factor used to account for offsets
between the \emph{NuSTAR} FPMA and FPMB detectors, again freezing it
at a value of 1 for the FPMB data and allowing it to vary from 0.9 to
1.1 for FPMA data. We also apply a constant factor to our
\emph{Chandra} data to similarly account for the offsets between
\emph{NuSTAR} and \emph{Chandra} data \citep{MADSEN2017}. We allow
this second constant to vary in a range of 0.1-10, given the
possibility of intrinsic source variability.

To model the spectrum for a given epoch or epochs, we loaded the
corresponding background model or spectral files created through the
processes detailed in \ref{sec:nubgd} and \ref{sec:chanbgd}. We
evaluated the following six spectral models: a powerlaw
(\texttt{zpowerlw}), a broken powerlaw (\texttt{zbknpower}), a double
broken powerlaw (\texttt{bkn2pow}), a log-parabola (\texttt{zlogpar}),
a cutoff powerlaw (\texttt{zcutoffpl}), and a thermal plasma model
(\texttt{apec}). For each model, we assumed a redshift of $z = 0.035$
and applied a photoelectric absorption model (\texttt{phabs}) with a
fixed Hydrogen column density value of nH =
$3.62\times10^{-20}\mathrm{cm}^{-2}$ \citep{H4PI2016}. For the thermal
plasma model, we additionally assumed a metallicity of 0.3 times solar
values.

We used the Bayesian X-ray Analysis (BXA) software from
\cite{BUCHNER2016}, which connects a nested sampling algorithm
\citep[UltraNest;][]{BUCHNER2021}) with the Python version of XSPEC,
pyXspec \citep{ARNAUD2021}. We found this to be the most effective and
stable method of spectral fitting. BXA also enables us to easily
compare non-nested models using the Bayes' evidence (Z), the
likelihood integrated over the parameter space \citep{KNUTH2014}. The
tested models reflect the possibility of X-ray emission from the
hotspot could potentially arise from inverse-Compton scattering,
synchrotron radiation, SSC, thermal emission, or any combination of
these processes.

\begin{table}
  \caption{Variability Analysis Data (\emph{NuSTAR}:3-78 keV, \emph{Chandra}: 0.5-8 keV) }
\centering
 \begin{tabular}{ll|ccccc}

Inst. & Proj. Code & Src. Counts & Bkg. Counts  & Src. Exposure & Bkg. Exposure & ECF \\ 
 &  & (ph) & (ph)  & ($10^{6}$ cm$^{2}\cdot$s) & ($10^{6}$ cm$^{2}\cdot$s) &  \\
\hline
\hline
        \emph{NuSTAR} & 60101047002 & 2127 & 1304 & 64.4 & - & 0.677 \\
        & 60701064002 & 3172 & 1510 & 109 & - & 0.671 \\
        & 60701064004 & 952 & 441 & 34.2 & - & 0.670 \\
        & 60701064006 & 3531 & 1638 & 123 & - & 0.670 \\
        &&&&&& \\
        \emph{Chandra} & 346 & 2969 & 160 & 11.9 & 12.1 & 0.973 \\
        & 3090 & 3229 & 156 & 12.5 & 12.3 & 0.982 \\
        & 4369 & 3573 & 359 & 13.7 & 13.5 & 0.980 \\
        & 11586 & 1105 & 104 & 3.97 & 3.89 & 0.979 \\
        & 12039 & 1890 & 212 & 7.13 & 7.00 & 0.982 \\
        & 12040 & 1411 & 148 & 5.07 & 5.06 & 0.977 \\
        & 14221 & 2781 & 203 & 9.90 & 8.63 & 0.981 \\
        & 15580 & 825 & 54 & 2.86  & 2.49 & 0.978 \\
        & 14222 & 3090 & 256 & 11.2 & 11.9 & 0.980 \\
        & 14223 & 2380 & 194 & 7.88 & 8.42 & 0.981 \\
        & 16478 & 1350 & 148 & 5.25 & 5.99 & 0.984 \\
        & 17574 & 1014 & 115 & 3.61 & 4.26 & 0.976 \\
        \hline
        \label{tab:vardata1}
        % \enddata
   \end{tabular}
\end{table}

We defined our prior distributions such that the analysis spans the
entire parameter space. In particular, we defined the priors for our
spectral parameters (e.g. $\Gamma$) and energy parameters to sample a
uniform distribution across all physically reasonable values. The
uninformative priors are thought to minimize bias in the posterior
distribution. We report the relevant prior distributions for all the
free parameters of the evaluated models in Table
\ref{tab:priors}. This table includes the model name, the XSPEC model
name, the model formula, the XSPEC parameter names, the type of prior
for the parameter (either uniform or log-uniform), and the minimum and
maximum value of the prior distribution.

\subsection{Variability analysis} \label{sec:var}

In previous work, \cite{MEYER2023} introduced a maximum-likelihood
based statistical test to analyze a series of CCD observations of
point sources for variability where the null hypothesis is a
non-variable Poisson source rate and a variable background rate. The
test uses a maximum likelihood model to return a p-value for a set of
single-region observations under the null hypothesis. Previously, this
method was used to conduct a variability survey of the large-scale,
X-ray jet population using \emph{Chandra} archival data of 53 sources,
including Pictor A \citep{MEYER2023}. While \emph{Chandra}'s high
angular resolution make it an ideal partner for general tests of
<resolved jet variability, the test is generally applicable to
e.g. other observatories with CCD detectors. Here we applied the
<variability test to both our \emph{Chandra} and \emph{NuSTAR} data on
Pictor A.

The maximum likelihood test is described in full in
\cite{MEYER2023}. Here, we will only briefly describe the necessary
elements of the test. For each observation entered into our analysis,
we required the following data: (1) the total number of photon counts
in the source region, (2) the number of background photons in the
source region, (3) the exposure in the source region, and (4) the
encircled counts fraction (ECF). For the analysis of the
\emph{Chandra} alone, we extracted this data for an energy range of
0.4-8 keV, and for \emph{NuSTAR} alone, we extracted it for the full
3-78 keV energy range. We also separately extracted the 3-8 keV counts
for a combined analysis using both observatories. We list the values
for this data for all observations used in the variability analysis
are given in Tables \ref{tab:vardata1} and \ref{tab:vardata2}. Using
this data, the maximum likelihood model returns an estimate of the
source flux, $\bar{\mu}$, under the steady-rate assumption and also
the probability of observing the data under the null hypothesis. A low
\emph{p}-value indicates that the null hypothesis can be rejected and
that the intrinsic source rate is therefore likely to be variable.

\begin{table}
  \caption{Variability Analysis Data (3-8 keV) }
\centering
 \begin{tabular}{ll|ccccc}
Inst. & Proj. Code & Src. Counts & Bkg. Counts  & Src. Exposure & Bkg. Exposure & ECF \\ 
 &  & (ph) & (ph)  & ($10^{6}$ cm$^{2}\cdot$s) & ($10^{6}$ cm$^{2}\cdot$s) & \\
\hline
\hline
        \emph{NuSTAR} & 60101047002 & 978 & 426 & 21.6 & - & 0.734 \\
        & 60701064002 & 1412 & 304 & 35.5 & - & 0.745 \\
        & 60701064004 & 396 & 89 & 11.0 & - & 0.736 \\
        & 60701064006 & 1497 & 333 & 40.3 & - & 0.746 \\
        &&&&&& \\
        \emph{Chandra} & 346 & 203 & 81 & 6.16 & 6.08 & 0.930 \\
        & 3090 & 281 & 56 & 7.82 & 8.11 & 0.956 \\
        & 4369 & 318 & 63 & 8.66 & 8.97 & 0.950 \\
        & 11586 & 106 & 42 & 2.86 & 2.88 & 0.884 \\
        & 12039 & 197 & 115 & 5.11 & 5.17 & 0.936 \\
        & 12040 & 150 & 86 & 3.63 & 3.68 & 0.911 \\
        & 14221 & 324 & 107 & 4.19 & 4.18 & 0.953 \\
        & 15580 & 390 & 134 & 2.31  & 2.37 & 0.932 \\
        & 14222 & 327 & 103 & 10.6 & 9.47 & 0.953 \\
        & 14223 & 115 & 30 & 8.61 & 9.02 & 0.948 \\
        & 16478 & 183 & 71 & 5.71 & 4.88 & 0.965 \\
        & 17574 & 128 & 63 & 4.19 & 3.55 & 0.942 \\
        \hline
 \end{tabular}
        \label{tab:vardata2}
%\enddata
\end{table}

We extracted the number of photon counts in our source region using
the \texttt{CIAO} tool \texttt{dmextract} and the total exposure for
the region using the same method on the exposure maps generated by
\emph{NuSTARDAS} \texttt{nuexpomap} for our \emph{NuSTAR} data and
\texttt{CIAO} tool \texttt{mkexpmap} for our \emph{Chandra} data. To
estimate (2) for \emph{Chandra}, we similarly extract the counts and
exposures of a corresponding background region and weight the
background counts by the ratio of the background exposure to the
source exposure. Unlike with \emph{Chandra}, \emph{NuSTAR}'s
background is spatially asymmetric across detectors, so we could not
use a background region separate from our source region. Instead, we
set the background exposure to be the same as our source exposure, and
used \texttt{nuskybgd} produced background maps to extract the
background counts within our source region.

In general for both \emph{NuSTAR} and \emph{Chandra} the PSF (and
therefore, the ECF) changes depending on the distance from the source
to the optical axis. We use our extracted regions to measure the mean
ECF for each observation. For \emph{Chandra}, we generated simulated
PSFs using the \texttt{CIAO} tool \texttt{simulate\_psf} and
\texttt{MARX} v5.4.0 and calculated the percentage of photons enclosed
within the source region. For \emph{NuSTAR}, we used \emph{NuSTARDAS}
product files to map the path of the optical axis for each observation
and took the median x- and y-coordinate of its path. We then simulated
the \emph{NuSTAR} PSF based on the distance between the source and the
optical axis and used this to calculate the ECF as was done for the
\emph{Chandra} observations.

\subsection{Image deconvolution with LIRA} \label{deconv}
To detect the inner jet or any features in the western hotspot, for example the ``bar'' seen in \emph{Chandra}~images, possibly buried in the raw NuSTAR observations, we deconvolved the merged sky image of Pictor A (see Figure \ref{fig:nu764X}) with an effective point spread function (PSF).

\begin{figure*}[t]
   \gridline{
   \fig{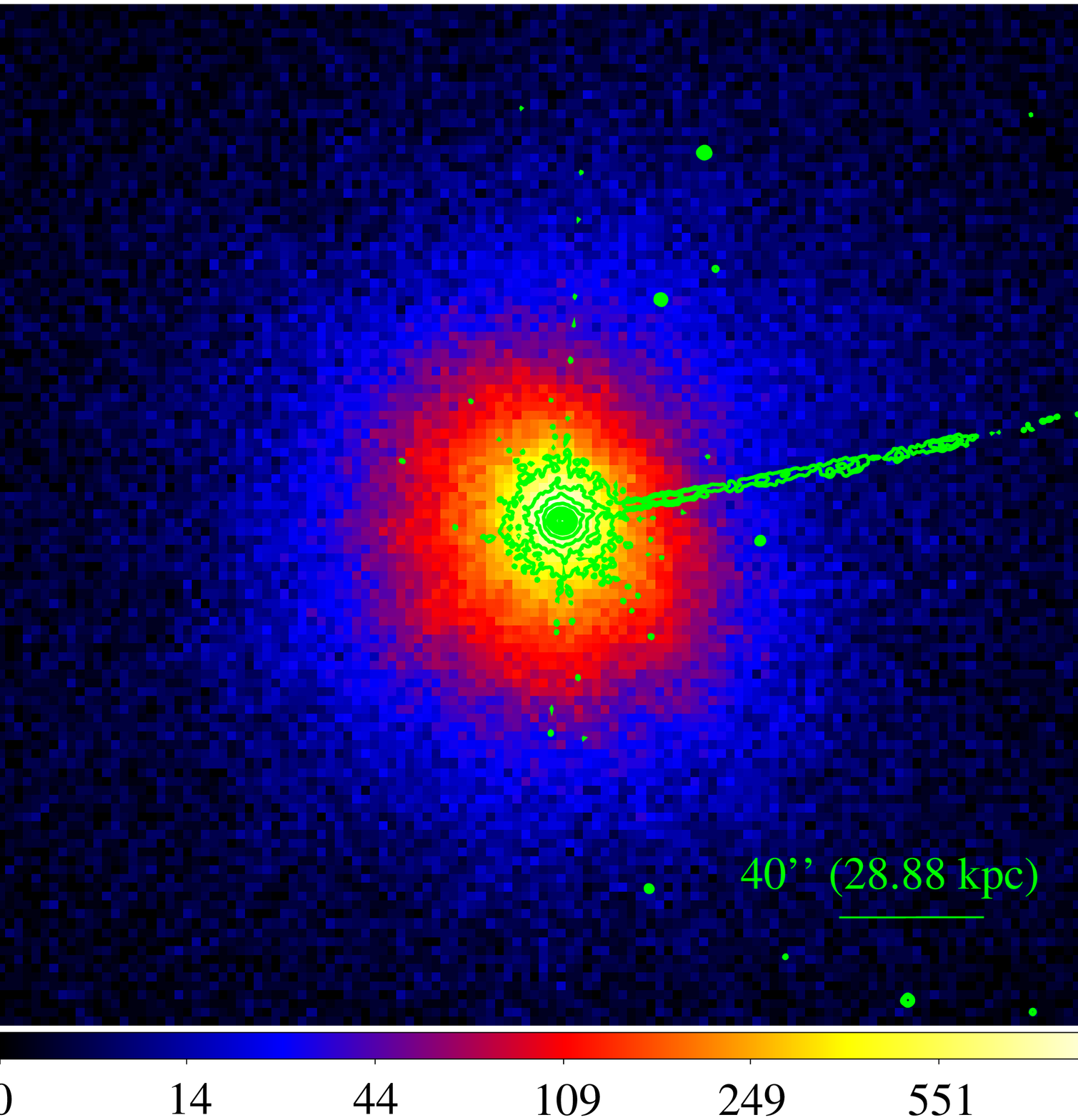}{0.25\textwidth}{(a) Core + Jet (3-12 keV)}
   \fig{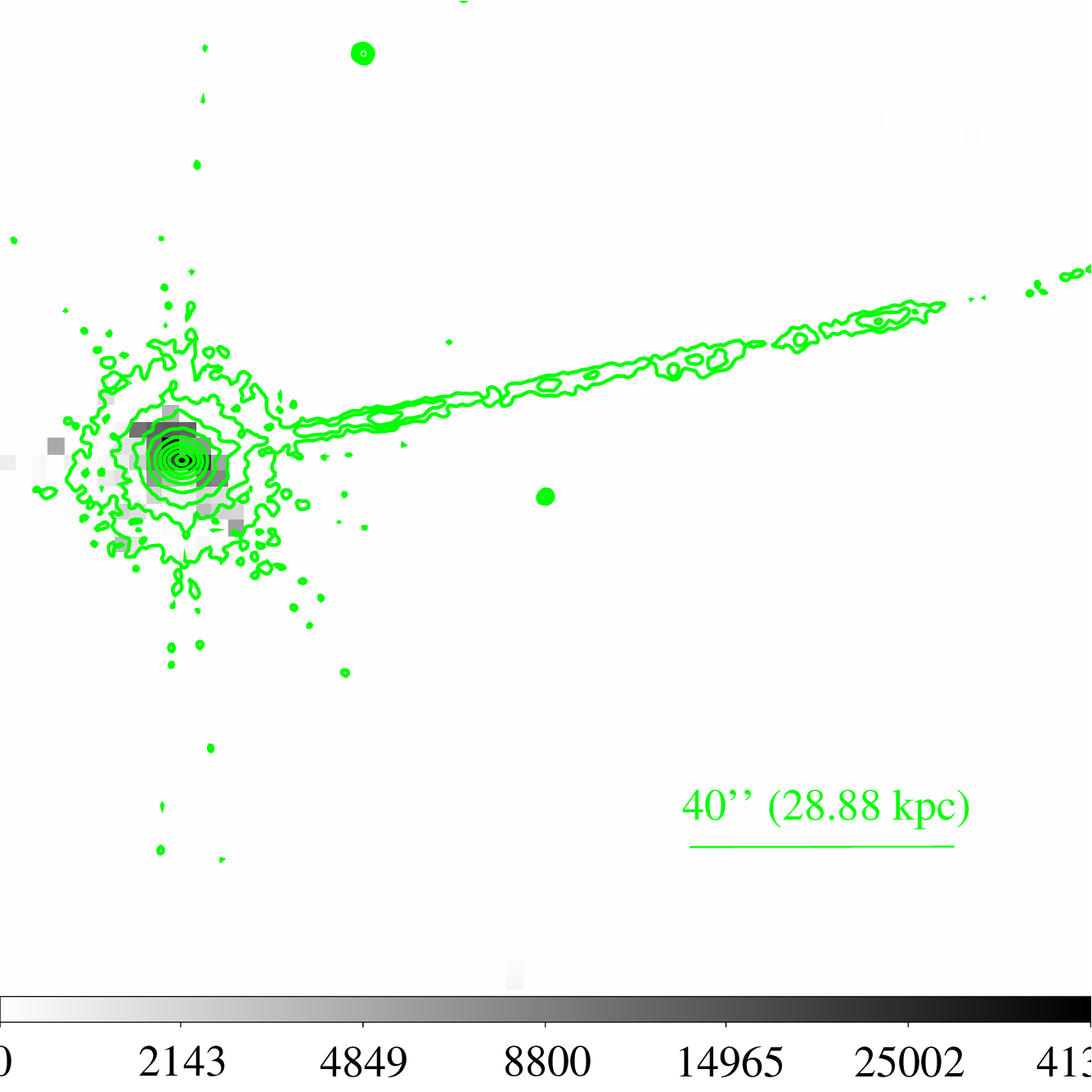}{0.25\textwidth}{(b) Deconvolved Core + Jet}
      \fig{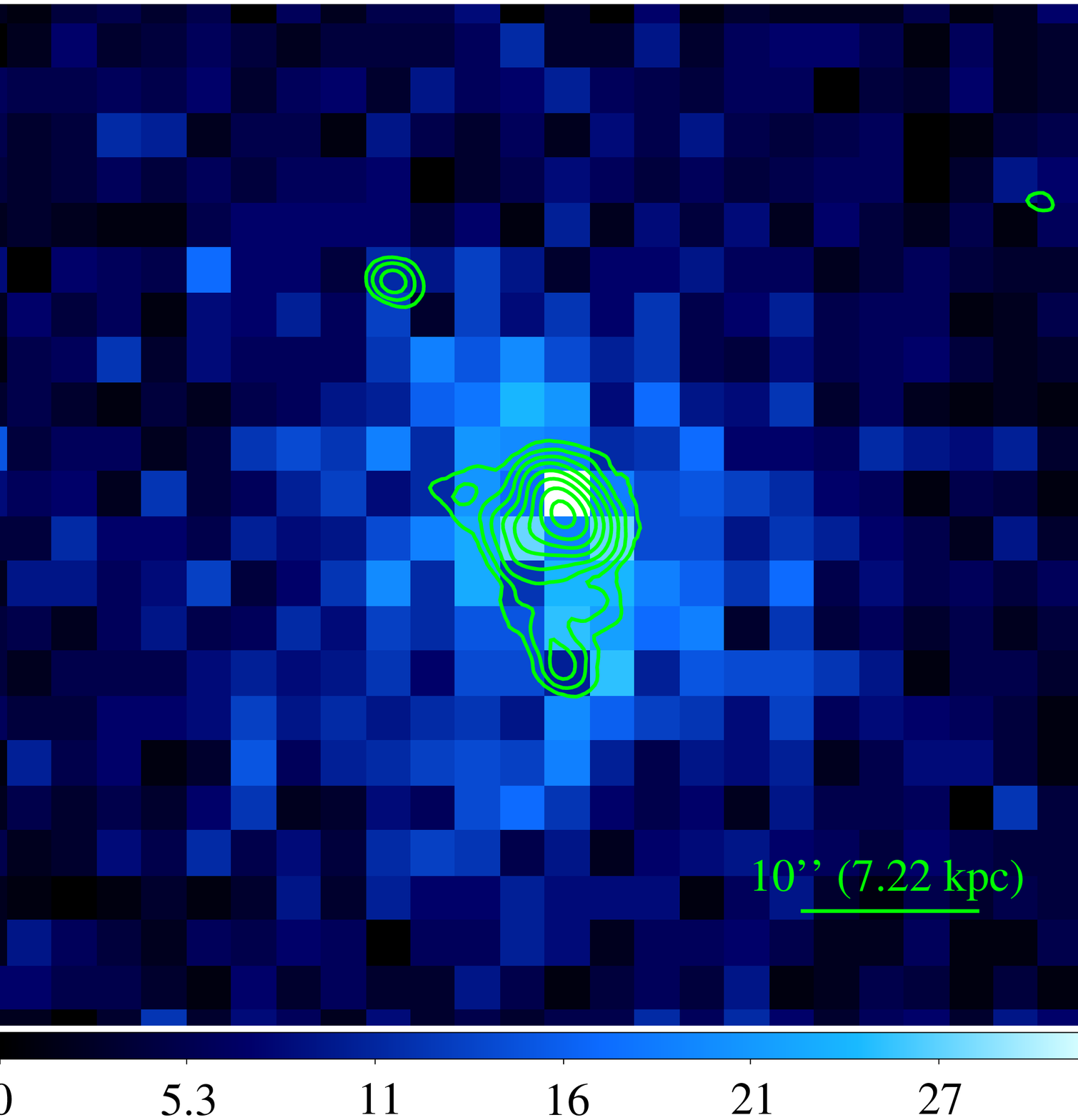}{0.25\textwidth}{(c) WHS}
   \fig{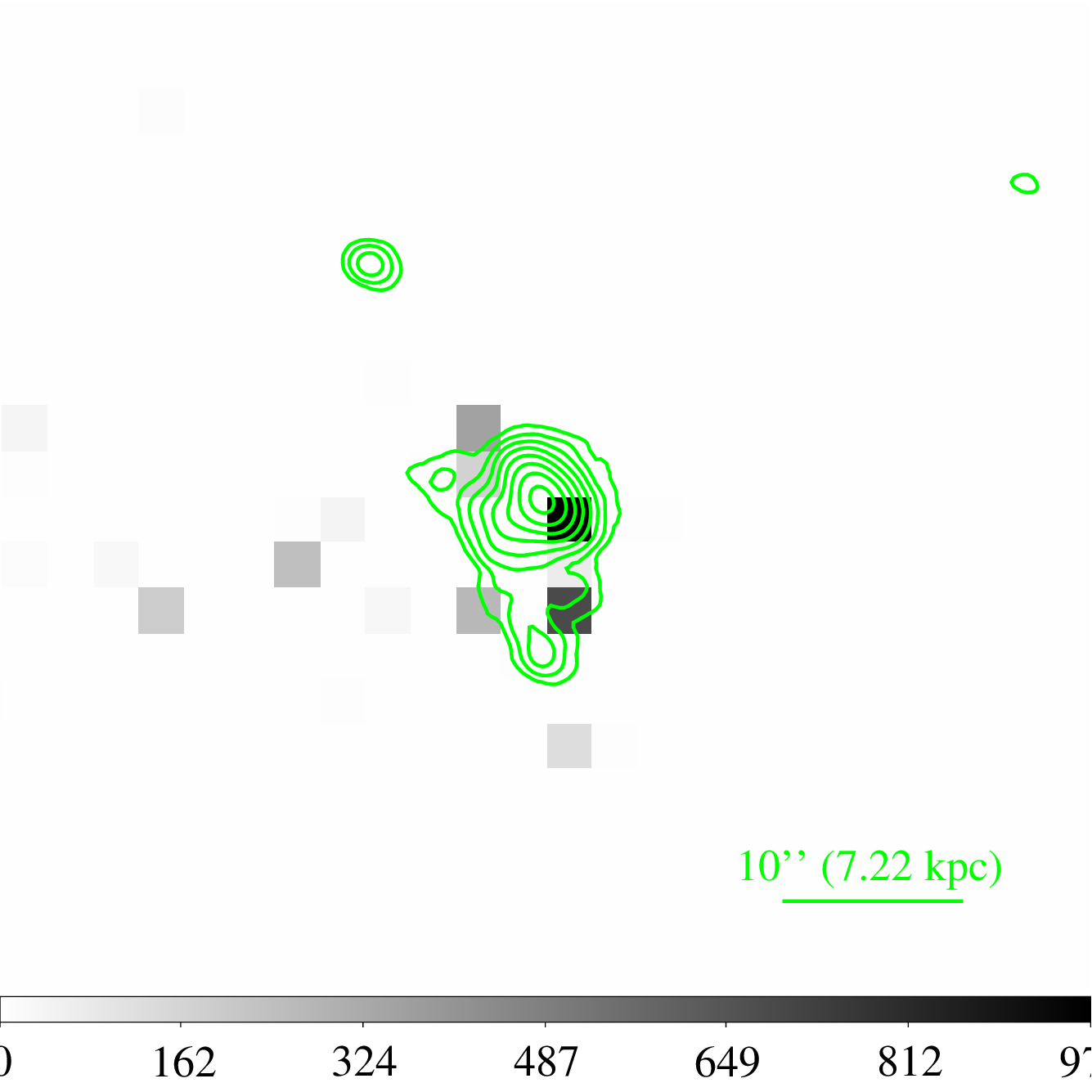}{0.25\textwidth}{(d) Deconvolved WHS}
   }

   \gridline{
   \fig{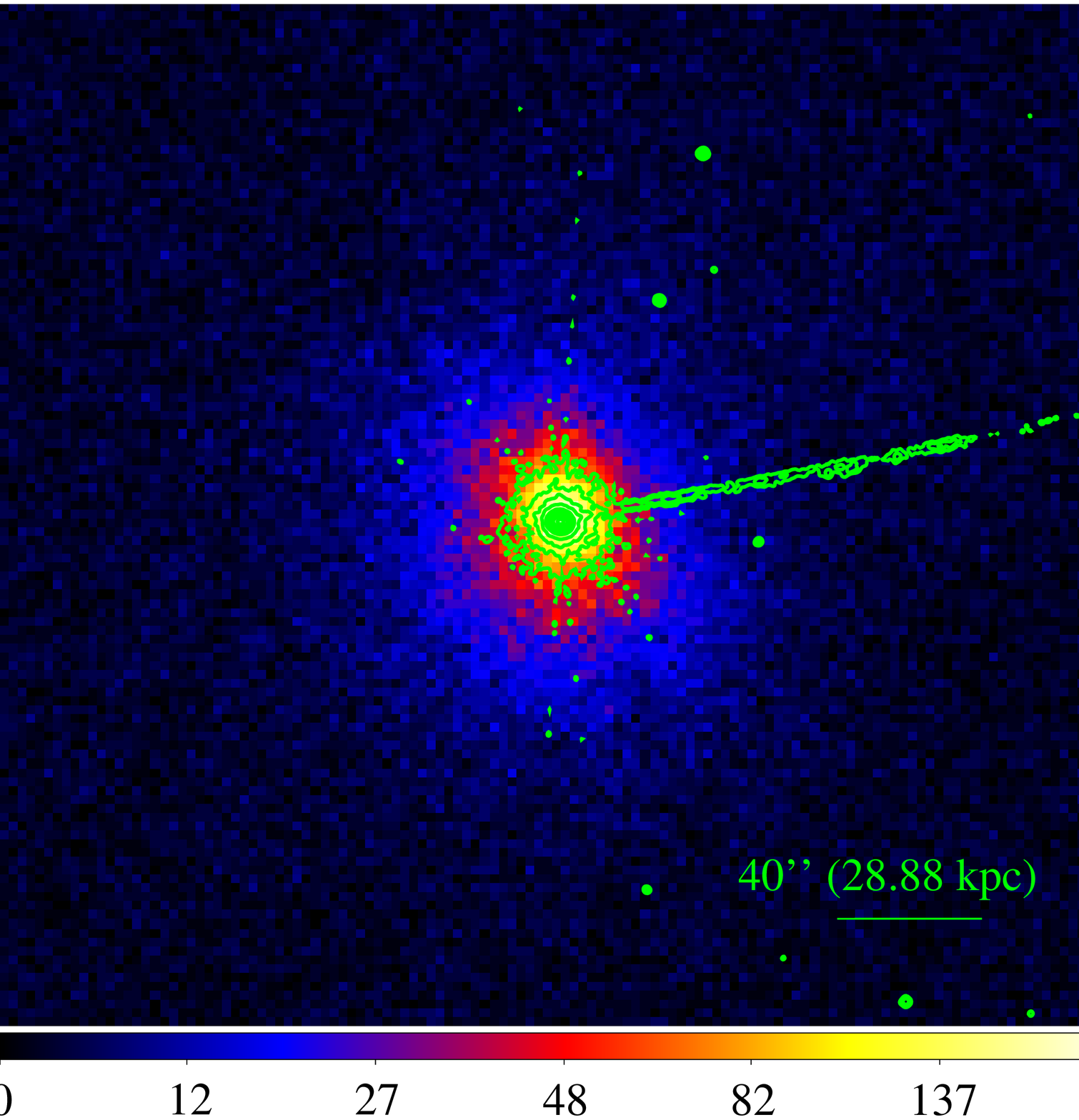}{0.25\textwidth}{(e) Core + Jet (12-78 keV)}
   \fig{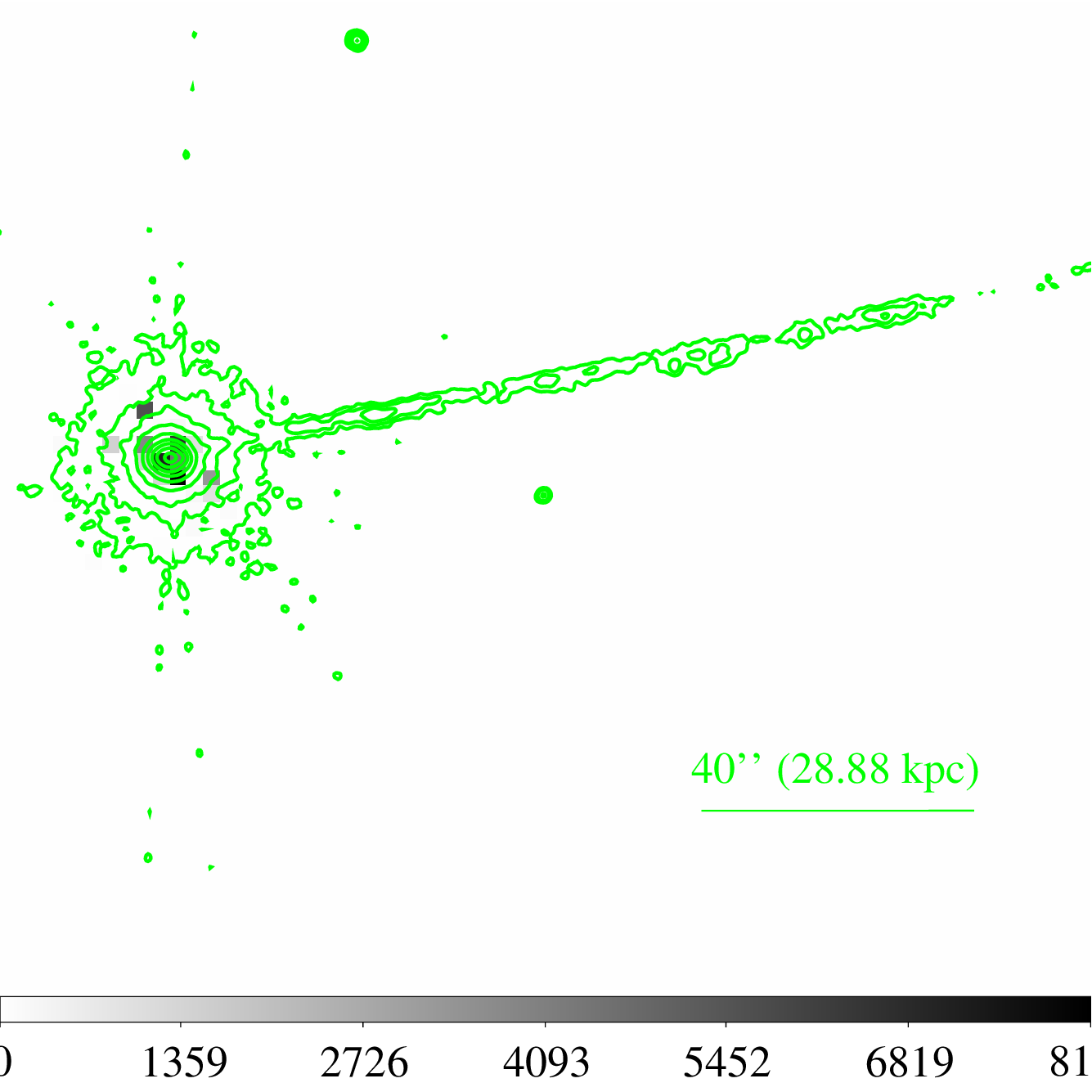}{0.25\textwidth}{(f) Deconvolved Core + Jet}
   \fig{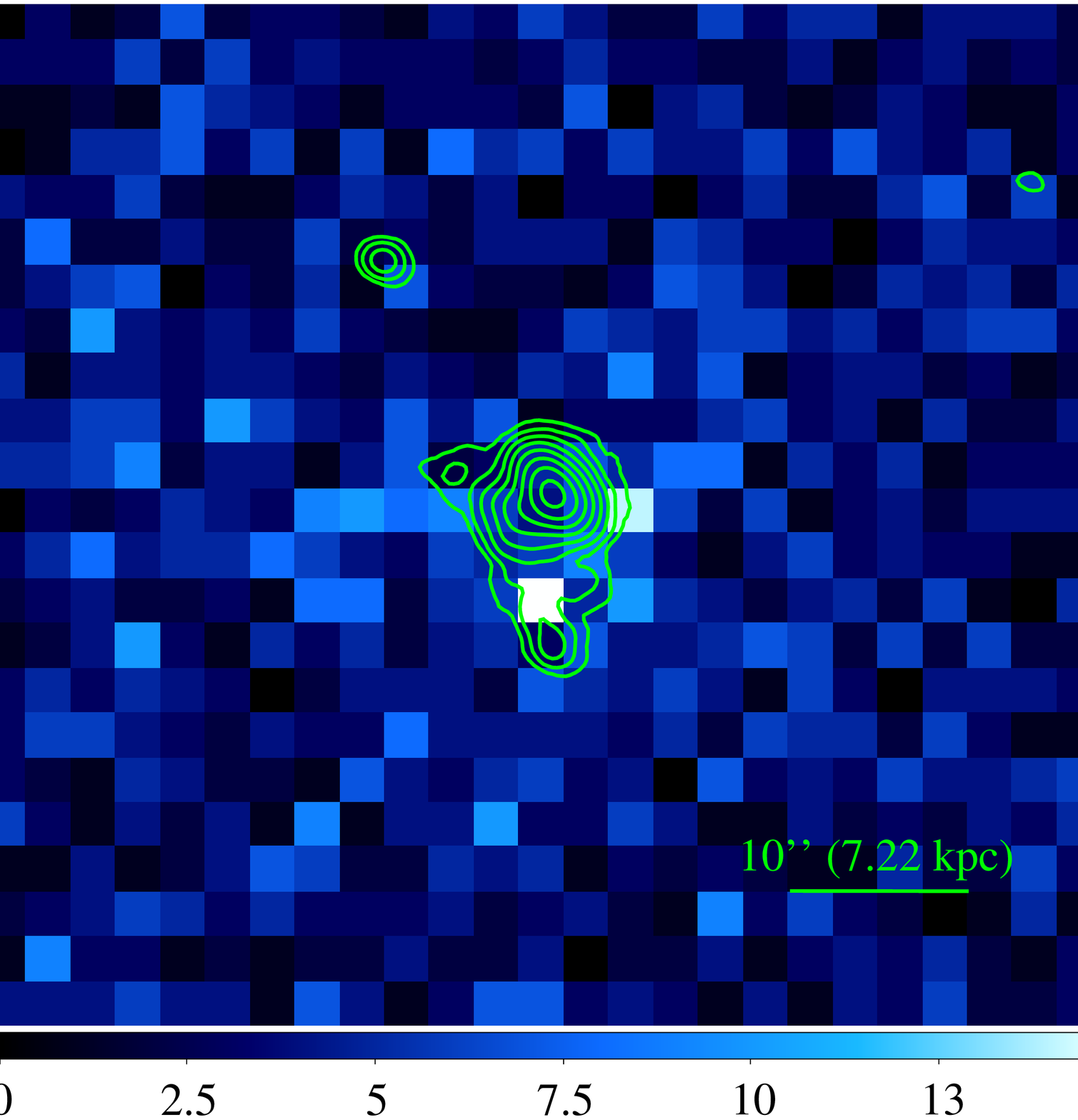}{0.25\textwidth}{(e) WHS }
   \fig{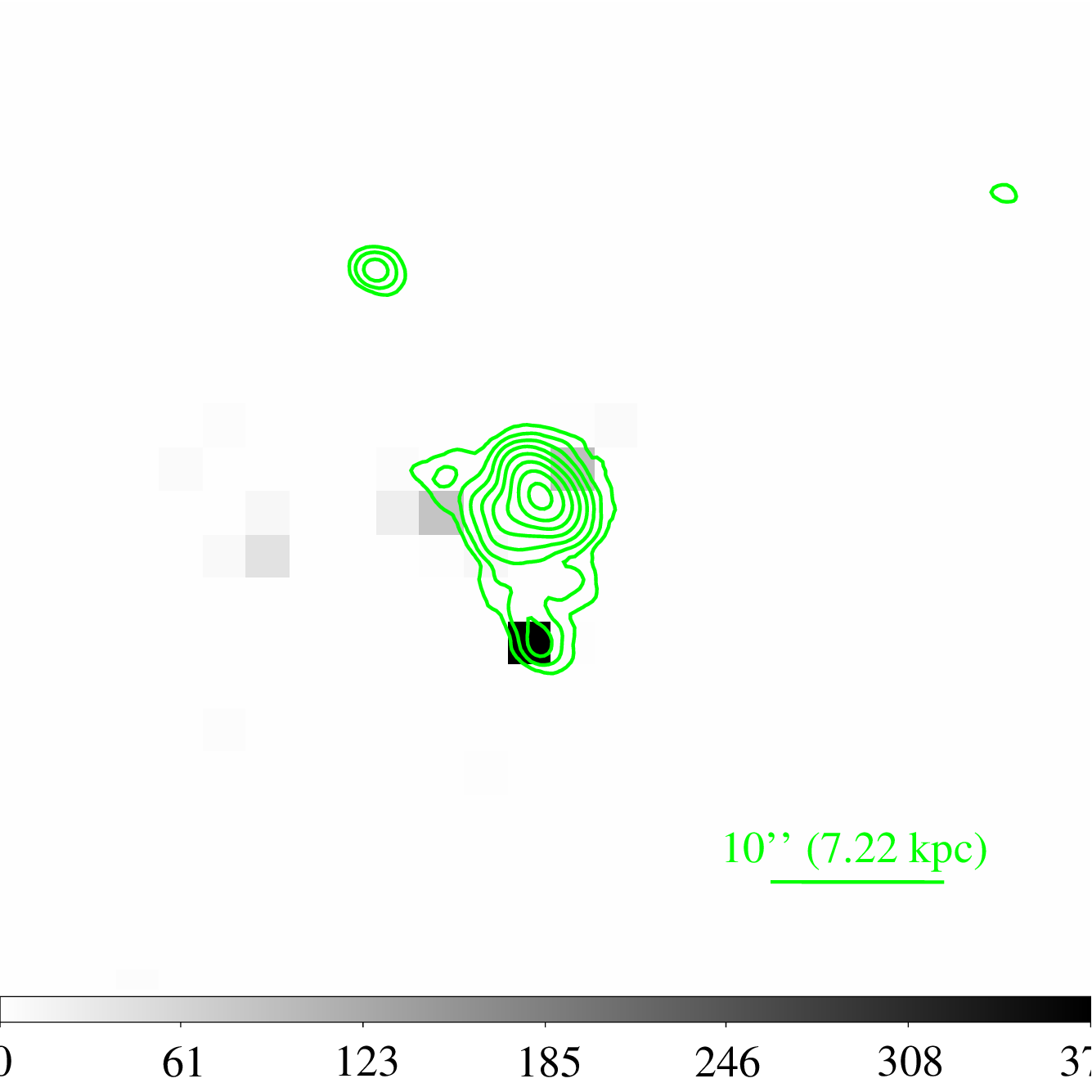}{0.25\textwidth}{(h) Deconvolved WHS}
   }
    \caption{LIRA-deconvolved images of Pictor A with the \emph{Chandra} contours overlaid on the top. (a) shows the deconvolved image of the core region where no evidence for an inner jet is found indicating its emission is within the error limits of the PSF of the core and background. (b) shows the deconvolved image of the western hotspot. The observed structure is roughly consistent with the \emph{Chandra} observations.}
    \label{fig:A_B_merged}
\end{figure*}

%Following \citep{2018ApJ...855...71S}, we used cross-correlation to correct astrometric offsets between individual observations. An image of the core for each observation was made with a 130\arcsec$\times$137\arcsec~rectangular region and the A module image with ObsID 60701064006 was chosen as the reference for its highest counts in the core.  Each remaining observation (from both the modules) was cross-correlated with the reference using \texttt{acrosscor} routine in CIAO v4.13 \citep{2006SPIE.6270E..1VF} and was fitted with a 2D Lorentzian profile in SHERPA  \citep{2001SPIE.4477...76F} to measure the offset between them. The offsets were applied to the sky images using \texttt{wcs\_update} routine with a root mean square correction of $(\Delta x,\Delta y)=(1.72\arcsec,3.32\arcsec)$. The corrected images were co-added to produce the  final merged image, shown in Figure \ref{fig:A_B_merged}. From the left, each panel shows an image with events falling in the range 3-78 keV, 3-12 keV and 12-78 keV, respectively.

We use a statistical tool called Low Count Image Reconstruction and
Analysis \citep[LIRA, ][]{esch2004image,stein2015detecting} to
deconvolve images. Although LIRA was originally designed to detect the
statistical significance of structures in low-count high-energy
observations, it also has been used to deconvolve sky images
\citep[e.g.,][]{Kashyap2017XrayingTE,2021ApJS..253...37R,
  Reddy_2023}. We refer the reader to \cite{stein2015detecting} for a
full statistical treatment. Briefly, LIRA models the brightness
distribution of an observation using two model images. One is a
baseline with known sources of emission (e.g., core and background),
which is specified by the user, and an added model, which is to be
inferred containing any emission over the baseline, required to
explain the observation. We can reconstruct the observation by
convolving their sum with the PSF. LIRA performs Bayesian analysis
using a flexible multi-scale representation of the observation to
infer the posterior distribution of the added model and samples it
with a Markov Chain Monte Carlo (MCMC) sampler to generate a series of
images. These images represent the observation's brightness
distribution after subtracting the baseline. With
\texttt{nuskybgd}-generated background as the baseline, we generate
2000 images after discarding the first 1000 images as \emph{burn-in}
and average them to produce a deconvolved image.

Due to the spatial dependence of NuSTAR's PSF, using a single PSF to
deconvolve both the core and the hotspot, which are separated by
$\sim$4', may lead to artifacts in the output images. Hence, we
separately deconvolve the core and the hotspot regions. Furthermore,
to determine the energy dependence of the structure in the jet and the
hotspot, we divide the image into two energy ranges of 3-12 keV and
12-78 keV, as shown in the top and bottom rows of Figure
\ref{fig:A_B_merged}. The left and right two panels in each row show
core and WHS images, respectively. We set the core image size to
128x128 and WHS to 64x64, following the input requirements of LIRA.

The top and bottom rows of panels in Figure \ref{fig:A_B_merged} show
the sky and LIRA-deconvolved images of the core and the WHS in the
energy ranges 3-12 keV and 12-78 keV, respectively. We find no
evidence for an inner jet in the core's deconvolution in both the
energy ranges (panels b and f), indicating its emission is within the
error limits of the core's PSF and background. This absence of
emission is expected in the sense that the flux from the inner jet is
close to the detection limit of NuSTAR. The 0.5-7 keV energy flux of
the brightest feature in the inner jet is $\approx2.5\times
10^{-14}$~erg cm$^{-2}$~s$^{-1}$, while the 3$\sigma$-sensitivity of
NuSTAR with a 1 Ms exposure\footnote{Measured in the 6-10 keV band}
(about three times larger than our combined 300 ks exposure) is
$2\times 10^{-15}$~erg
cm$^{-2}$~s$^{-1}$~\citep{Harrison_2013}. Finally, because we also
find no evidence for a jet beyond 12 keV (Fig.
\ref{fig:A_B_merged}(e)), we use the estimated 10-30 keV sensitivity
of $10^{-14}$~erg cm$^{-2}$~s$^{-1}$ as an upper limit for each
knot. Hence, we take $4\times10^{-14}$~erg cm$^{-2}$~s$^{-1}$~as an
upper limit on the inner jet's 10-30 keV flux.

Panels (d) and (f) show the LIRA-deconvolved images of the WHS in the
3-12 keV and 12-78 keV bands, respectively, where the detected
structure appears spatially correlated with previously known
features. The 3-12 keV images show point sources at the southern end
of the bar and the hotspot, consistent with the structure found in
\textit{Chandra} images. Interestingly, the 12-78 keV band lacks a
point source near the hotspot, but we find a bright point feature
spatially coincident with the bar's southern peak. This feature
indicates the jet presumably produces an electron population capable
of generating hard X-rays when it enters the turbulent hotspot
region. However, the limited statistics in the bar region preclude a
detailed spectral analysis and require a much deeper observation to
understand the true nature of this hard X-ray emission.

\section{Results}

We present the fit results for our analyses in Tables
\ref{tab:nustarlastresults} (\emph{NuSTAR} epochs 1, 2, and all epochs
combined) and \ref{tab:lastresults} (\emph{Chandra} all epochs
combined and \emph{NuSTAR}+\emph{Chandra} combined fit). The results
for the individual \emph{Chandra} epochs are given in Appendix Table
\ref{tab:chandralastresults}.  In these tables we give the best-fit
values and $1\sigma$ posterior distribution of the model parameters
detailed in Table \ref{tab:priors} for each data set as calculated
through our BXA analysis. We report only on the most likely models for
a given epoch or combined fit as determined by the Bayesian evidence
(see below). \textcolor{black}{In all but one case, the powerlaw (PL),
  broken powerlaw (Bkn. PL), and double broken powerlaw (Dbl. Bkn. PL)
  were the three preferred models according to this metric (refer to
  Table \ref{tab:priors} for exact definitions of these models). We
  report energy parameters ($E_{brk}, E_{brk,1}, E_{brk,2}$) in units
  of keV, normalizations ($N_{1}$) in units of nJy, and photon indices
  ($\Gamma, \Gamma_{1}, \Gamma_{2}, \Gamma_{3}$).}

\begin{sidewaystable}
  \caption{\emph{NuSTAR} X-ray Spectral Fitting Results }
  \centering
  \footnotesize
\begin{tabular}{llllllBll|ccc}
NS-1 & \textcolor{black}{Powerlaw (PL)} & $N_{1} = 84.6^{+19.2}_{-12.4}$ & $\Gamma = 1.94^{+0.10}_{-0.09}$ & - & - & - & - & -0.12 & 0.271 & 37.8/49 = 0.771 \\
 & \textcolor{black}{Broken powerlaw (Bkn. PL)} & $N_{1} = 92.9^{+22.2}_{-17.5}$ & $\Gamma_{1} = 2.00^{+0.16}_{-0.15}$ & $E_{brk} = 4.03^{+0.65}_{-0.71}$ & $\Gamma_{2} = 1.93^{+0.12}_{-0.11}$ & - & - & -0.07 & 0.304 & 37.0/47 = 0.796 \\
 & \textcolor{black}{Double broken powerlaw* (Dbl. Bkn. PL*)} & $N_{1} = 93.9^{+21.1}_{-18.2}$ & $\Gamma_{1} = 2.01^{+0.15}_{-0.16}$ & $\Gamma_{2} = 1.94^{+0.17}_{-0.13}$ & $E_{brk,1} = 3.98^{+0.72}_{-0.65}$ & $E_{brk,2} = 7.54^{+1.75}_{-1.66}$ & $\Gamma_{3} = 1.96^{+0.18}_{-0.14}$ & 0.00 & 0.357 & 36.7/45 = 0.823 \\
NS-2 & PL* & $N_{1} = 126.9^{+12.7}_{-9.9}$ & $\Gamma = 2.05^{+0.05}_{-0.04}$ & - & - & - & - & 0.00 & 0.338 & 142.9/159 = 0.899 \\
 & Bkn. PL & $N_{1} = 127.2^{+22.2}_{-21.8}$ & $\Gamma_{1} = 2.06^{+0.11}_{-0.13}$ & $E_{brk} = 5.40^{+1.79}_{-1.65}$ & $\Gamma_{2} = 2.05^{+0.07}_{-0.07}$ & - & - & -0.03 & 0.315 & 141.6/157 = 0.902 \\
 & Dbl. Bkn. PL & $N_{1} = 125.8^{+24.7}_{-25.4}$ & $\Gamma_{1} = 2.05^{+0.14}_{-0.18}$ & $\Gamma_{2} = 2.05^{+0.10}_{-0.10}$ & $E_{brk,1} = 4.03^{+0.68}_{-0.66}$ & $E_{brk,2} = 7.49^{+1.75}_{-1.65}$ & $\Gamma_{3} = 2.04^{+0.09}_{-0.10}$ & -0.01 & 0.330 & 141.7/155 = 0.914 \\
NS-All & PL* & $N_{1} = 118.9^{+8.7}_{-9.2}$ & $\Gamma = 2.03^{+0.04}_{-0.04}$ & - & - & - & - & 0.00 & 0.373 & 196.9/211 = 0.933 \\
 & Bkn. PL & $N_{1} = 120.4^{+24.0}_{-26.0}$ & $\Gamma_{1} = 2.04^{+0.14}_{-0.19}$ & $E_{brk} = 3.97^{+0.72}_{-0.71}$ & $\Gamma_{2} = 2.02^{+0.05}_{-0.05}$ & - & - & -0.08 & 0.310 & 196.0/209 = 0.938 \\
 & Dbl. Bkn. PL & $N_{1} = 119.1^{+25.2}_{-24.0}$ & $\Gamma_{1} = 2.04^{+0.14}_{-0.17}$ & $\Gamma_{2} = 2.02^{+0.10}_{-0.10}$ & $E_{brk,1} = 4.05^{+0.65}_{-0.73}$ & $E_{brk,2} = 7.38^{+1.88}_{-1.69}$ & $\Gamma_{3} = 2.03^{+0.09}_{-0.09}$ & -0.08 & 0.310 & 201.5/207 = 0.974 \\
\hline
\end{tabular}
\tablecomments{   * denotes the preferred model, according to the Bayesian evidence}
\tablecomments{\textcolor{black}{Break energies ($E_{brk}, E_{brk,1}, E_{brk,2}$) are reported in units of keV; 1 keV flux normalizations ($N_{1-kev}$) are reported in units of nJy; photon indices ($\Gamma, \Gamma_{1}, \Gamma_{2}, \Gamma_{3}$) are dimensionless}}
\tablecomments{\textcolor{black}{Refer to Table \ref{tab:priors} for exact definitions of the models and their respective parameters.}}

 % \enddata
 \label{tab:nustarlastresults}
%\tablecomments{   * denotes the preferred model, according to the Bayesian evidence}
\end{sidewaystable}

We compared the alternative models based on three different metrics
which are suitable for comparing non-nested models. The first is the
Bayesian evidence, defined as the integral over the parameter space:
$Z = \int
\pi(\vv{\theta})\mathrm{exp}[-\frac{1}{2}C(\vv{\theta})]d\vv{\theta}$,
where $\vv{\theta}$ is the parameter vector. The second is the
Bayesian information criterion, $BIC = C - n\times ln(d)$
\citep{SCHWARZ1978}, where $C$ is the log-likelihood value, $d$ is the
number of data points, and $n$ is the number of degrees of
freedom. The BIC is an approximation under the assumption that the
posterior distribution is strongly single-peaked, which is
well-supported by the posterior distributions for our data. The final
comparison statistic we use is the Akaike information criterion, $AIC
= C - 2\times n$ \citep{AKAIKE1974}. The $AIC$ is a more conservative
metric which measures the information loss of a specific model and is
useful for determining if a model has the appropriate number of
degrees of freedom. In all cases included in this study, comparing
models using $\Delta AIC$ and $\Delta BIC$ only reinforces the
paradigm established by the Bayesian evidence. As such, we omit these
values from Tables \ref{tab:nustarlastresults} and
\ref{tab:lastresults}. Although we emphasize that it is not useful for
comparison nor is it used in fitting at any point, we do calculate the
$\chi^{2}_{red} = \chi^{2}/\mathrm{DOF}$ for each model as a check of
the goodness-of-fit. This value is calculated by loading the best-fit
model from the BXA analysis into \texttt{XSPEC} and binning the
spectra to at least 30 counts per bin using \texttt{ftgrouppha}. The
Bayesian evidence, the relative probabilities of the different models
(which are calculated directly from the Bayesian evidence), and the
$\chi^{2}_{red}$ are all recorded in Tables
\ref{tab:nustarlastresults} and \ref{tab:lastresults}.

%\begin{splitdeluxetable*}{llllllBll|ccc}
%\tablecaption{\emph{Chandra} and \emph{Chandra+NuSTAR} X-ray Spectral Fitting Results \label{tab:lastresults}}
%\tablehead{\colhead{Inst.-Epoch} & \colhead{Model} & \colhead{$N_{1-\mathrm{keV}}$ (nJy)} & \colhead{param. 1} & \colhead{param. 2} & \colhead{param. 3 (keV)} & \colhead{param. 4} & \colhead{param. 5} & \colhead{log(Z)} & \colhead{p(M|D)} & \colhead{$\chi^2$/dof} }
%\colnumbers
%\startdata
\begin{sidewaystable}
  \caption{\emph{Chandra} and \emph{Chandra+NuSTAR} X-ray Spectral Fitting Results}
\centering
 \begin{tabular}{llllllBll|ccc}
Inst.-Epoch & Model & $N_{1-\mathrm{keV}}$ (nJy) & param. 1 & param. 2& param. 3 (keV) & param. 4 & param. 5 & log(Z) & p(M|D) &$\chi^2$/dof\\
   \hline
   \hline
CH-All & PL & $N_{1} = 95.1^{+0.7}_{-0.8}$ & $\Gamma = 1.98^{+0.01}_{-0.01}$ & - & - & - & - & $-0.66$ & 0.113 & 464.0/598 = 0.776 \\
 & Bkn. PL & $N_{1} = 95.0^{+0.7}_{-0.7}$ & $\Gamma_{1} = 1.96^{+0.02}_{-0.02}$ & $E_{brk} = 2.98^{+0.70}_{-0.51}$ & $\Gamma_{2} = 2.15^{+0.10}_{-0.09}$ & - & - & -0.14 & 0.372 & 462.6/596 = 0.776 \\
 & Dbl. Bkn. PL* & $N_{1} = 95.0^{+0.7}_{-0.7}$ & $\Gamma_{1} = 1.95^{+0.02}_{-0.02}$ & $\Gamma_{2} = 2.20^{+0.16}_{-0.14}$ & $E_{brk,1} = 3.04^{+0.42}_{-0.51}$ & $E_{brk,2} = 4.87^{+1.67}_{-0.88}$ & $\Gamma_{3} = 1.92^{+0.23}_{-0.22}$ & 0.00 & 0.515 & 453.9/594 = 0.764 \\
CH+NU-All & PL & $N_{1} = 120.1^{+3.5}_{-3.8}$ & $\Gamma = 2.01^{+0.01}_{-0.01}$ & - & - & - & - & $-3.36$ & $4.36\times10^{-4}$ & 660.9/802 = 0.824 \\
 & Bkn. PL & $N_{1} = 119.9^{+4.0}_{-4.2}$ & $\Gamma_{1} = 2.01^{+0.01}_{-0.01}$ & $E_{brk} = 13.65^{+4.36}_{-9.26}$ & $\Gamma_{2} = 2.00^{+0.15}_{-0.19}$ & - & - & $-3.38$ & $4.16\times10^{-4}$ & 462.6/596 = 0.776 \\
 & Dbl. Bkn. PL* & $N_{1} = 97.8^{+5.0}_{-4.4}$ & $\Gamma_{1} = 2.49^{+0.01}_{-0.11}$ & $\Gamma_{2} = 1.98^{+0.02}_{-0.03}$ & $E_{brk,1} = 0.69^{+0.04}_{-0.03}$ & $E_{brk,2} = 11.44^{+6.08}_{-8.23}$ & $\Gamma_{3} = 2.05^{+0.09}_{-0.18}$ & 0.00 & 0.999 & 453.9/594 = 0.764 \\
 & PL+Exp. Cutoff & $N_{1} = 105.4^{+9.7}_{-9.5}$ & $\Gamma = 1.90^{+0.05}_{-0.06}$ & - & $E_{cut} = 57.53^{+14.6}_{-16.8}$ & - & - & - & - & - \\
%\enddata
\hline
\end{tabular}
 \tablecomments{   * denotes the preferred model, according to the Bayesian evidence}
 \label{tab:lastresults}
 \end{sidewaystable}
%\end{splitdeluxetable*}

\subsection{\emph{NuSTAR} spectral fitting}

According to the Bayesian evidence resulting from our analysis of the
combined \emph{NuSTAR} data (recorded in Table
\ref{tab:nustarlastresults}), the simple power-law is the slightly
preferred model for the 3-20 keV band. (The broken power law and
double broken power law appear at first to be acceptable, if slightly
less-preferred models, but see discussion of parameter values below.)
The exponential cutoff powerlaw, log-parabola, and thermal models are
effectively ruled out by the evidence. The simple power law model is
also favored when comparing the $\Delta AIC$ and $\Delta BIC$ for the
different models.

The best-fit folded power law model and posterior range are plotted
alongside the \emph{NuSTAR} FPMA and FPMB data in the top panel of
Figure \ref{fig:nu_spec}, and the residuals of the model are plotted
in the bottom panel. This simple power law fit of the 3-20 keV
spectrum has a best-fit photon index of $\Gamma = 2.03\pm0.04$ and a
normalization of $118.9^{+8.7}_{-9.2}$ nJy at 1 keV. The maximum
likelihood value for the constant offset factor we introduce to
account for sensitivity differences between the FPMA and FPMB modules
is $f = 1.01\pm0.03$ which is in agreement with previous estimates of
a $\pm5\%$ calibration difference between the two detectors
\citep{MADSEN2015}.

{\centering
\begin{figure}[th]
    %\centeringl
    \gridline{
     \fig{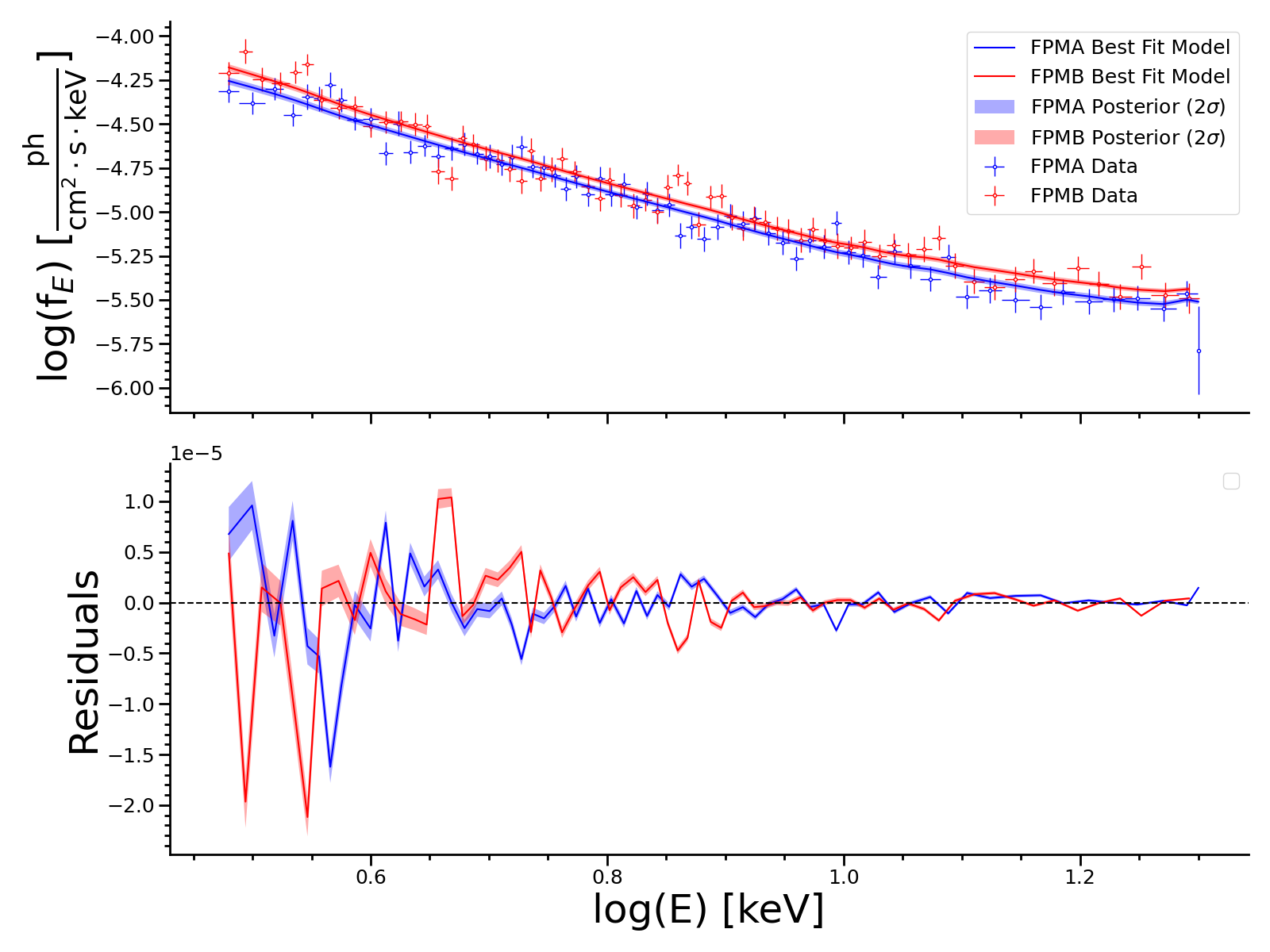}{0.45\textwidth}{}
    }
    \caption{(Top) Photon flux density versus energy for the combined
      NuSTAR epochs, shown separately for the FPMA (blue) and FPMB
      (red) modules. For plotting purposes only, the data are
      collected into 66 and 69 spectral bins for the FPMA and FPMB
      detectors respectively across an energy range of 3-20 keV. Each
      point is the averaged photon flux as calculated using the source
      counts and exposure for that spectral bin. The solid line plots
      the flux density of the folded best-fit powerlaw model including
      the effects of galactic absorption. The shaded contours around
      either line displays the 2$\sigma$ range in flux density of the
      posterior distribution. (Bottom) The residuals of the best-fit
      power law model and posterior distribution as compared to the
      spectral data. As with the top plot, the solid lines plots the
      residuals between the best-fit model and the data and the shaded
      contours plot the 2$\sigma$ range of the posterior distribution
      for both the FPMA and FPMB modules. }
    \label{fig:nu_spec}
\end{figure}}

We find no evidence of a spectral break or cutoff in our \emph{NuSTAR}
X-ray data alone. The maximum likelihood parameter values for the
broken powerlaw and double broken powerlaw models are both asymptotic
approximations of the simple power law. That is, the posterior
distribution of the broken powerlaw is strongly peaked at a parameter
distribution such that $\Gamma_{1}\sim\Gamma_{2}$ and the posterior
distribution of the double broken powerlaw is strongly peaked such
that $\Gamma_{1}\sim\Gamma_{2}\sim\Gamma_{3}$.

We also fit the 2 epochs of \emph{NuSTAR} data separately. The
Bayesian evidence of the second epoch (which has triple the exposure
of the first) again favored the simple powerlaw model, with the
best-fit broken powerlaw and double broken powerlaw models being
slightly less probable and showing little change in spectral index
across the energy band. For the first epoch, the double broken
powerlaw is actually the slightly preferred model, however the
parameter best-fit values and errors make it clear that it is
effectively equivalent to a single constant power-law spectrum, with
$\Gamma_{1} = 2.01^{+0.15}_{-0.16}$, $\Gamma_{2} =
1.94^{+0.17}_{-0.13}$, and $\Gamma_{3} = 1.96^{+0.18}_{-0.14}$.

Assuming a simple powerlaw as the correct model, the photon index
showed no significant variation between the two major epochs, ranging
from $\Gamma = 1.94^{+0.10}_{-0.09}$ for epoch 1 to $\Gamma =
2.05^{+0.05}_{-0.04}$ for epoch 2. However, the best-fit normalization
changed substantially between the two epochs, with an approximately
$\sim50\%$ increase in value from epoch 1 to epoch 2, suggesting
variability in the hard X-ray flux. This is further examined and
quantified in Section ~\ref{sec:var2}.

\subsection{\emph{Chandra X-ray spectrum}}

According to the Bayesian evidence resulting from our analysis of the
combined \emph{Chandra} data (recorded in Table
\ref{tab:lastresults}), the double broken power-law is the slightly
preferred model for the 0.5-7 keV band (although see discussion
below). The broken power law and simple power law are also acceptable
if slightly less-preferred models, while the exponential cutoff
powerlaw, log-parabola, and thermal models are effectively ruled out
by the evidence.

Interestingly, when we fit the individual epochs of \emph{Chandra}
data separately, the simple powerlaw (1, 3-5) and broken powerlaw (2,
6, 7) are consistently favored over the double broken powerlaw (albeit
only slightly). The preference for the double broken powerlaw in the
combined fit might be the result of variations from epoch to epoch
which manifest as an additional spectral break in the combined data
set. Because of this, and the weak constraints on the second break
energy and third photon index, we instead opt to consider the broken
power law model as the ``true" spectral model as it is only slightly
less preferred than the double broken powerlaw in the combined
dataset.

The best-fit broken powerlaw model has a primary photon index of
$\Gamma_{1} = 1.95\pm0.02$, a secondary photon index of $\Gamma_{2} =
2.15^{+0.10}_{-0.09}$, a break energy at $E_{brk} =
2.98^{+0.70}_{-0.51}$, and a normalization of $N_{1} = 95.0\pm0.7$ nJy
at 1 keV. This folded model and its posterior range are plotted
alongside the combined 0.5-7 keV \emph{Chandra} data in the top panel
of Figure \ref{fig:chan_spec}, and the residuals of the model are
plotted in the bottom panel. These results are in rough agreement with
\cite{HARDCASTLE2016}, whose analysis of the \emph{Chandra}
observations of the Western hotspot also slightly preferred the broken
power law model fit to the spectrum. The location of the break energy
at just below the \emph{NuSTAR} band combined with the preferred
simple powerlaw model for the hard X-ray spectrum of the hotspot also
provides more evidence in support of a singly broken powerlaw.

{\centering
\begin{figure}[th]%\centeringl
    \gridline{
     \fig{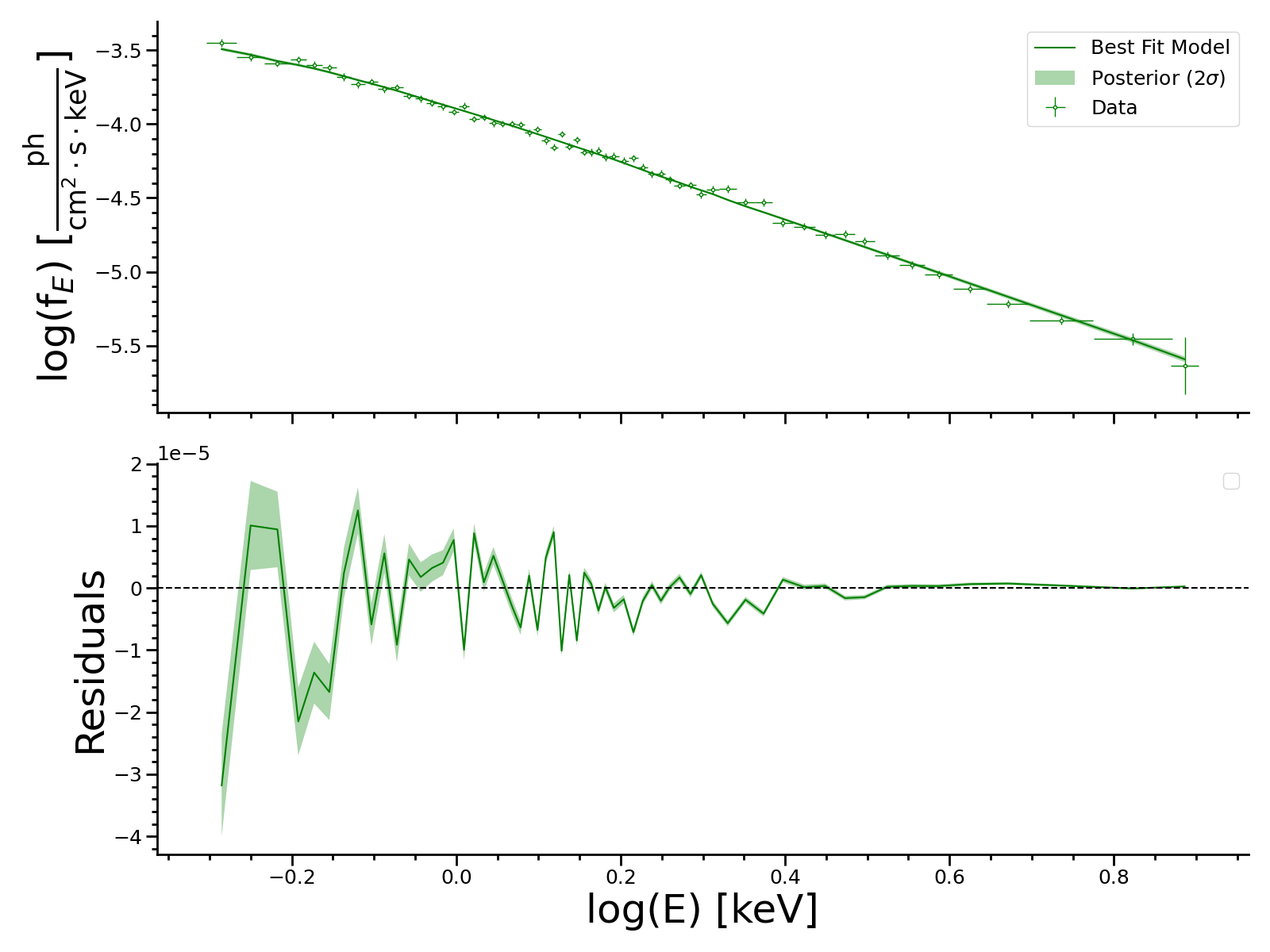}{0.45\textwidth}{}
    }
    \caption{(Top) Photon flux versus energy for the combined
      \emph{Chandra} data. For plotting purposes only, the data are
      collected into 187 spectral bins across an energy range of 0.5-7
      keV. Each point is the averaged spectrum across the 11
      observations for that spectral bin, calculated using the total
      counts and exposure for that spectral bin. The solid green line
      plots the flux of the best-fit broken power law model including
      the effects of galactic absorption. For visual purposes only,
      the background spectrum is subtracted. The shaded green contours
      around either line displays the 2$\sigma$ range in flux density
      of the posterior distribution. (Bottom) The residuals of the
      best-fit power law model and posterior distribution as compared
      to the spectral data. As with the top plot, the solid green line
      plots the residuals between the best-fit model and the data and
      the shaded green contours plot the 2$\sigma$ range of the
      posterior distribution.}
    \label{fig:chan_spec}
\end{figure}}

\begin{figure}[th]
    %\centeringl
    \gridline{
     \fig{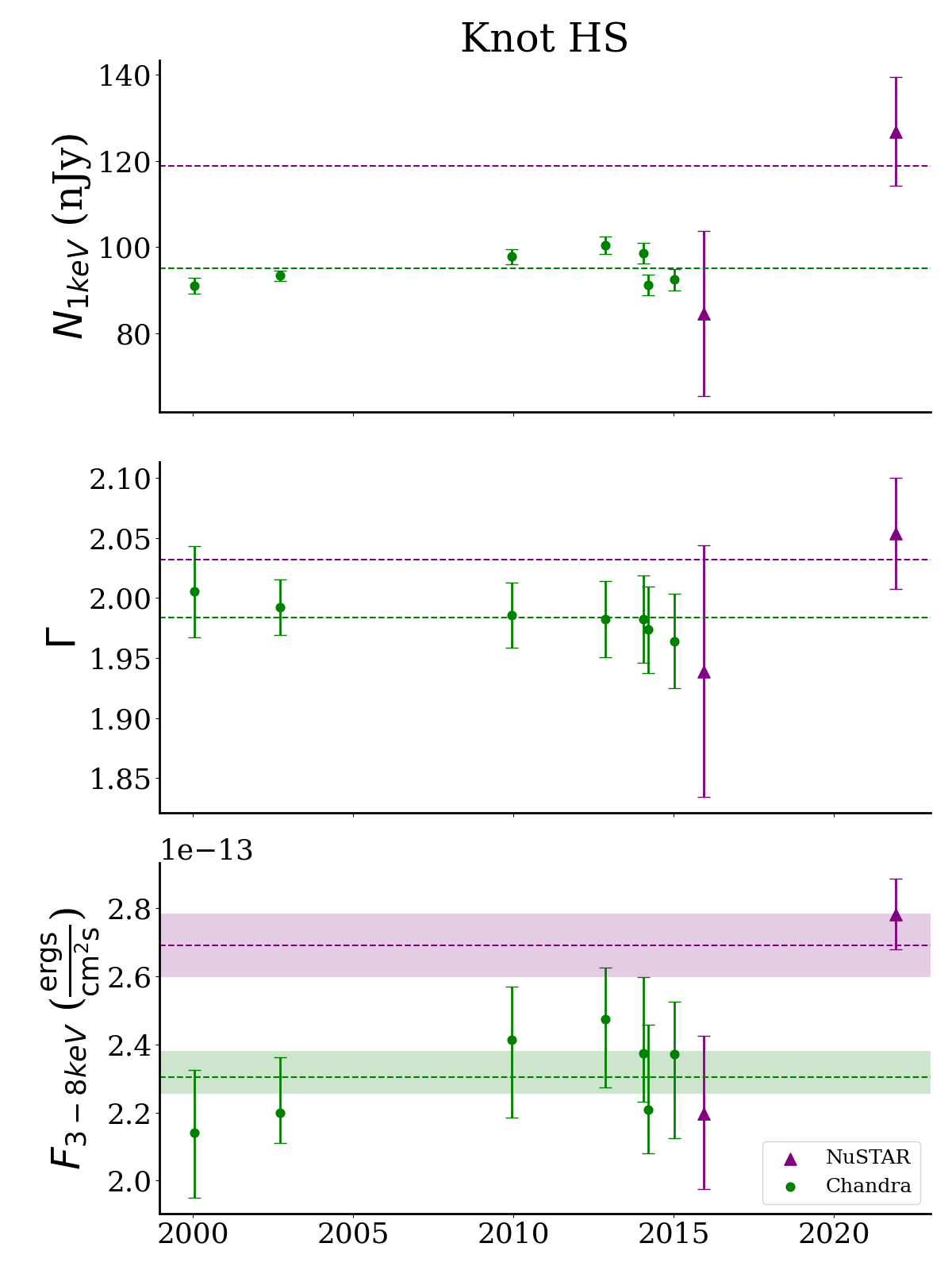}{0.5\textwidth}{} }
    \caption{A plot showing the evolution of the powerlaw photon
      normalization at 1 keV, $N_{1}$ (top), the powerlaw spectral
      index, $\Gamma$ (middle), and the integrated flux in the shared
      3-8 keV band (bottom) over time. The error bars for the first
      two are determined via the Bayesian analysis, while the errors
      for the bottom plot are calculated via the \texttt{XSPEC} cflux
      command. The best-fit values for the combined data of all epochs
      for textit{NuSTAR} FPMA (red) and FPMB (blue) and \emph{Chandra}
      (green) are plotted as dashed lines. The normalization shows
      slight changes over time with a significant offset between
      best-fit values between \emph{NuSTAR} and
      \emph{Chandra}. Despite large changes in the normalization
      between \emph{NuSTAR} epochs 1 and 2, the uncertainties due to
      lower exposure limit the statistical significance of our
      variability analysis. In contrast, $\Gamma$ shows little change
      over time and little variation between the different telescopes,
      maintaining a nearly constant spectrum ($\Gamma\sim2$) across
      all data. }
    \label{fig:gamma_norm}
\end{figure}

Similar to our \emph{NuSTAR} analysis, comparing the changes in the
best-fit single powerlaw models between various \emph{Chandra epochs}
reveals that the photon index varies little, ranging from $\Gamma =
1.93^{+0.06}_{-0.06}$ to $\Gamma = 2.01^{+0.04}_{-0.03}$. Also
similarly, albeit smaller in scale, we see a roughly 10$\%$ decrease
in the normalization from \emph{Chandra} epoch 4 to epoch 5, or a
2.33$\sigma$ change. This is similar to the variability previously
reported by \cite{HARDCASTLE2016}. The values of the normalization and
integrated flux in the 3-8 keV flux band shared by both \emph{NuSTAR}
and \emph{Chandra} across the different epochs of data from both
telescopes are plotted in Figure \ref{fig:gamma_norm}, with the photon
index in the lower panel. Further quantification and discussion of the
X-ray variability is continued in Section ~\ref{sec:var2}.

\subsection{Combined Chandra + NuSTAR X-ray spectrum}

According to the Bayesian evidence resulting from our analysis of the
combined fit of our \emph{Chandra}+\emph{NuSTAR} data (recorded in
Table \ref{tab:lastresults}), the double broken powerlaw is the
preferred model for the 0.5-20 keV wide band. However, similar to our
analysis of the two telescopes independently, a close inspection of
the parameter values reveals that this preference does not manifest as
a ``true" spectral break. The double broken powerlaw has an
exceptionally steep primary photon index which is not seen in previous
analyses ($\Gamma_{1} = 2.49^{+0.01}_{-0.11}$) and a low energy for
the primary spectral break ($E_{brk,1} = 0.69^{+0.04}_{-0.03}$). This
is likely a case of overfitting and attempting to correct for offsets
between the different datasets. The broken power law and simple power
law are also acceptable if less-preferred models, while the
exponential cutoff powerlaw, log-parabola, and thermal models are
effectively ruled out by the evidence.

The best-fit broken power law model and posterior range are plotted
alongside the combined \emph{Chandra} and \emph{NuSTAR} data in the
top panel of Figure \ref{fig:comb_spec}, and the residuals of the
model are plotted in the bottom panel. This broken power law model has
a best-fit photon indices of $\Gamma_{1} = 2.25^{+0.01}_{-0.03}$ and
$\Gamma_{2} = 1.98^{+0.01}_{-0.01}$, a break energy of $E_{brk} =
0.76^{+0.07}_{-0.06} \mathrm{keV}$, and a normalization of
$107.6^{+4.4}_{-3.8}$ nJy at 1 keV.

To constrain the possibility of a cutoff to the spectrum (which is of
interest for its physical implications for the maximum electron
energy, $E_{max}$; see Section \ref{sec:disc}) we fit the broadband
X-ray spectrum (3-78 keV) using a powerlaw with exponential cutoff,
finding a $1\sigma$lower limit of $E_{cut} > 40.7$ keV for the cutoff
energy. The best-fit parameter values for the overall fit are included
in Table \ref{tab:lastresults}, but since this model incorporates data
for the entire \emph{NuSTAR} spectrum, we do not attempt to compare to
the other models via information criterion or the Bayesian
evidence. (The same model fit to the 3-20 keV data range was
significantly disfavored compared to the other models.

Although the Bayesian evidence suggests that the broken powerlaw and
double broken powerlaw models are significantly more probable than the
simple powerlaw model, a closer examination of the results suggest
they must be interpreted with caution. Firstly, the best fit primary
break energy value for both the broken and double broken powerlaw
models is approximately $\sim0.75$ keV, close to the lower boundary of
the spectral range and differing from the value of $\sim$3 keV for the
Chandra fit alone. The photon index after the break is approximately
$\sim2$, which is in agreement with the best fit simple powerlaw
photon index calculated for most epochs of both \emph{NuSTAR} and
\emph{Chandra} data. In the case of the double broken powerlaw model,
the second spectral break shows no significant change in spectral
index for before and after the second break ($1.98\pm0.02$ to
$2.03\pm0.06$), and thus can be disregarded. Importantly, the joint
fit also includes a relative normalization factor for \emph{NuSTAR} to
force agreement with \emph{Chandra}; the value is $f\sim0.80\pm0.02$,
a larger difference than expected for instrumental effects alone. The
combination of evident variability (which also complicated results of
the \emph{Chandra} analysis presented in the last section)
%The slight upward shift of the \emph{NuSTAR} flux is also evident in the SED, as discussed below. 
{\centering
\begin{figure}[th]
    %\centeringl
    \gridline{
     \fig{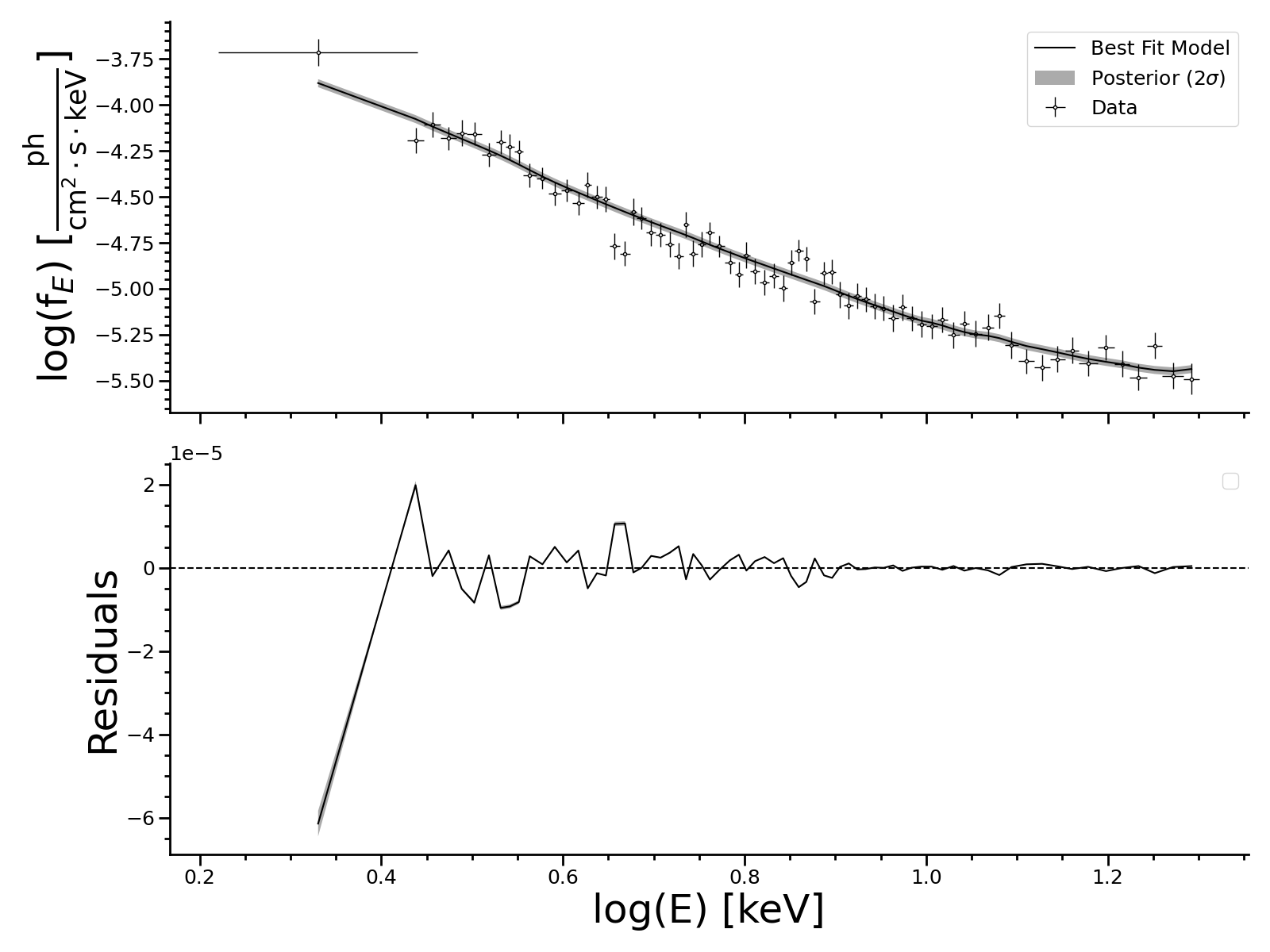}{0.45\textwidth}{}
    }
    \caption{(Top) The combined spectral data of all 11 \emph{Chandra}
      and 8 \emph{NuSTAR} FPMA and FPMB observations used in this
      study fit by a single model. The data are binned into 70
      spectral bins across an energy range of 0.5-20 keV for visual
      purposes only. The solid black line plots the flux of the
      best-fit broken power law model including the effects of
      galactic absorption and the contributions of background
      radiation. The shaded contours around the line displays the
      2$\sigma$ range in flux density of the posterior
      distribution. (Bottom) The residuals of the best-fit power law
      model and posterior distribution as compared to the spectral
      data. As with the top plot, the solid black line plots the
      best-fit residuals and the shaded contours plot the 2$\sigma$
      range of the posterior distribution. }
    \label{fig:comb_spec}
\end{figure}}

\subsection{Spectral energy distribution}
For comparison and to produce the most complete spectrum of Pictor A's
western hotspot to date, we collected multifrequency archival data
from previous studies, including radio and infrared observational data
from the VLA \citep{MEISENHEIMER1997}, observations from Spitzer
\citep{WERNER2012}, WISE \citep{ISOBE2017}, and SPIRE
\citep{ISOBE2020}, as well as additional archival data recorded in the
aforementioned studies.

The full spectral energy distribution (SED) of the western hotspot is
shown in Figure \ref{fig:sed}. We include the model of
\cite{ISOBE2020} to describe the low energy emission, consisting of a
primary cutoff powerlaw and an `excess' broken cutoff powerlaw to
account for the observed flux at mid- and far-infrared frequencies. We
additionally plot $\nu F_{nu}$ as calculated from the unfolded data of
our independent \emph{Chandra} and \emph{NuSTAR} datasets. The data
points are corrected for background, instrument response, and
absorption using \texttt{XSPEC}'s \texttt{plot ufspec} command.

There is an offset in normalization between our \emph{NuSTAR} and
\emph{Chandra} data which is evident both in the SED and in the
best-fit parameter values recorded in Table \ref{tab:lastresults},
suggesting that the hotspot may have been in a somewhat higher state
during the recent \emph{NuSTAR} observation. Because the XSPEC model
by default normalizes the flux at 1 keV (technically below the
\emph{NuSTAR} sensitivity range), this apparent variability likely
explains the apparent shift of the spectral break energy in the
combined fit as an artifact of the forced agreement. A visual
inspection of the SED suggests that there is indeed a slight break at
around 2 keV. At higher energies, the spectral fits for essentially
all data sets here considered agree on a very nearly perfectly
constant spectrum with no other break or cutoff.

{\centering
\begin{figure*}[th]
    \gridline{
     \fig{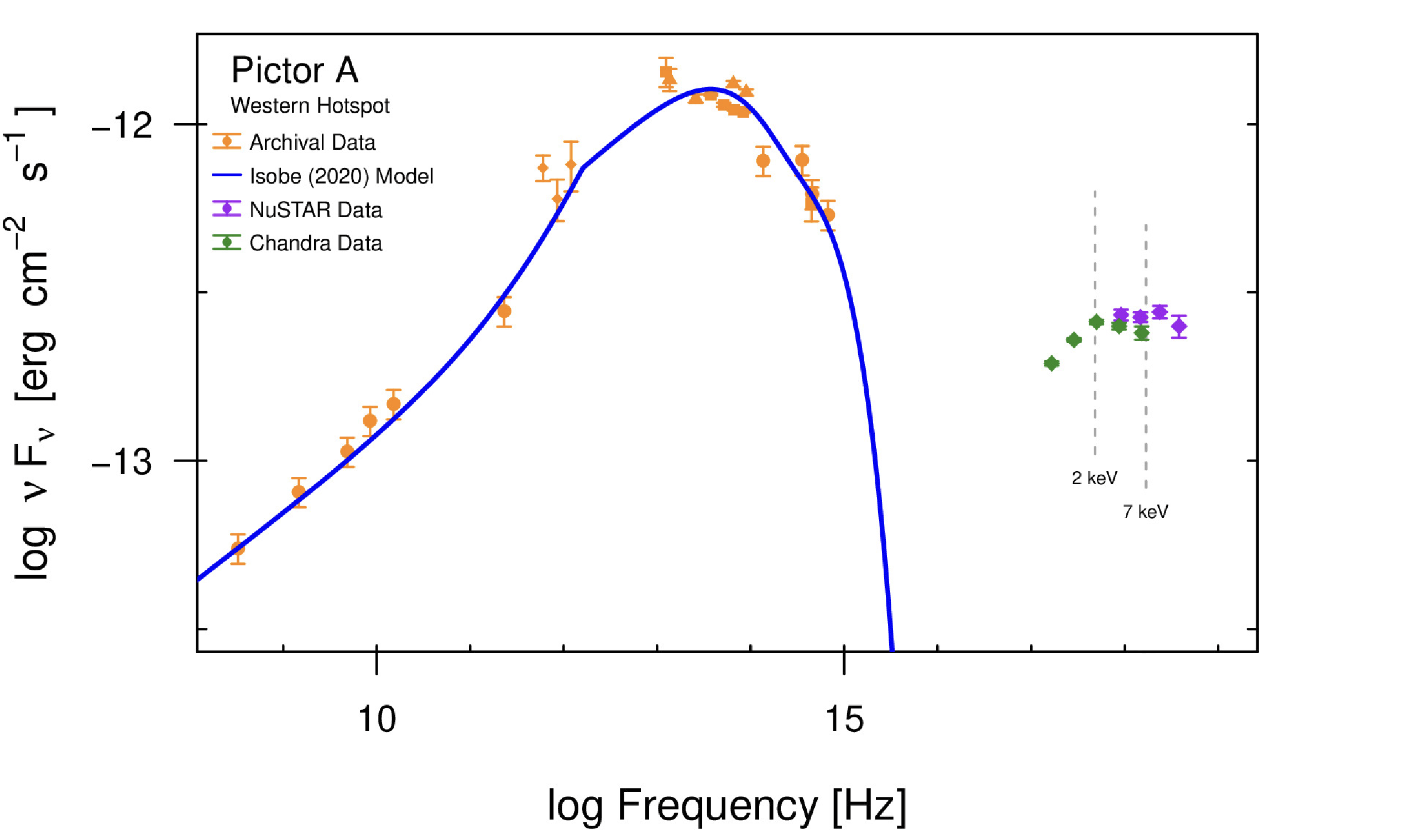}{0.9\textwidth}{}
     }
    \caption{The SED of the Western hotspot of Pictor A. Archival data
      taken from \cite{MEISENHEIMER1997} (circles), \cite{WERNER2012}
      (squares), \cite{ISOBE2017} (triangles), and \cite{ISOBE2020}
      (diamonds) are plotted in orange. This archival data is
      consistent with a synchrotron spectrum modeled as a cutoff
      powerlaw and a broken cutoff powerlaw to account for excess IR
      emission, thought to be due to synchrotron emission from
      discrete zones with enhanced magnetic field strength
      \citep{ISOBE2020}. We plot the unfolded X-ray spectral data
      points based on the best-fit models for \emph{NuSTAR} (purple)
      and \emph{Chandra} (green). We calculated the unfolded data
      points corrected for the instrument response and the galactic
      absorption using \texttt{XSPEC}'s \texttt{plot ufspec} command,
      which scales the data by a factor equal to the ratio of the
      unfolded model flux and the folded model flux. The model
      predictions from the two observatories shows a significant
      difference in the measured flux, which is further evidence of
      short timescale variability.}
    \label{fig:sed}
\end{figure*}}

\subsection{Variability analysis} \label{sec:var2}
We find strong evidence of statistically significant variability in
the total \emph{Chandra}-measured X-ray flux for the western
hotspot. When we include all 11 \emph{Chandra} observations, the
returned \emph{p}-value is $p = 2\times10^{-11}$, allowing us to
reject the steady rate hypothesis at a significance of $6.76\sigma$,
approximately twice the significance over previous reports. We also
analyzed smaller subsets of observations, in order to constrain the
shortest timescale of the flux variability. For example, applying our
variability analysis to only epochs 5 and 6 of the \emph{Chandra}
data, which are separated in time by approximately 2 months, still
gives a \emph{p}-value of $p = 9\times10^{-4}$, or $3.32\sigma$. This
suggests significant X-ray flux variations on the timescales of
months, which by light-crossing time arguments would place an upper
limit of $r<.01$ pc on the varying region. This result also supports
the findings of \cite{HARDCASTLE2016}, who reported a change in the
flux of the western hotspot of approximately $10\%$ on a timescale of
approximately $\sim$1 month.

For the variability analysis of the \emph{NuSTAR} data alone, we
recorded the necessary data outlined in Section \ref{sec:var}
separately for both the FPMA and FPMB detectors. We then generated a
combined data set, using the total photon counts, total exposure, and
average ECF from the two detectors, which are presented in Table
\ref{tab:vardata1}. We then analyzed this combined data set using our
maximum likelihood analysis, which returns $p = 0.082$, equivalent to
$1.74\sigma$.

The two epochs of NuSTAR data alone do not demonstrate highly
significant variability, but when we include the full 22 year
time-span of \emph{NuSTAR} and \emph{Chandra} in the shared 3-8 keV
energy band (data recorded in \ref{tab:vardata2}), our maximum
likelihood method returned a \emph{p}-value below the machine
precision value of $R$, $p < 2.2\times10^{-16}$, giving a significance
of at least $\sigma > 8.21$. This result for the Western hotspot is
the strongest evidence yet of short-timescale ($\sim$years)
variability across the X-ray hotspot population. These results, along
with additional tests for short-term variability are summarized in
Table \ref{tab:variability}, where we list the instrument(s), epochs,
resultsing p-values and equivalent significance, and the longest time
baseline of the data set.

\begin{table}
   \caption{Variability results} 
\centering
 \begin{tabular}{lc|ccc}  
  Instrument & \hspace{.75cm}Epochs\hspace{.5cm} & \hspace{.5cm}\emph{p}-value\hspace{.75cm} & \hspace{.33cm}Significance ($\sigma$)\hspace{.33cm}& \hspace{.33cm} ($\Delta$t)\hspace{.33cm}  \\
    \hline
    \hline
    \emph{Chandra} & 1-3 & 0.064 & 1.85 & 9 yr \\
    \emph{Chandra} & 5-6 & 8.68e-4 & 3.13 & 3 mo \\
    \emph{Chandra} & 4-7 & 3.19e-5 & 3.75 & 2 yr \\
    \emph{Chandra} & All & 3.33e-11 & 6.53 & 15 yr \\
    \emph{NuSTAR} & 2 & 0.820 & 0.23 & 1 wk \\
    \emph{NuSTAR} & All & 0.082 & 1.74 & 6 yr \\
    \emph{NuSTAR}+\emph{Chandra} & All & $<$2.2e-16 & $>$8.21 & 22 yr\\
    \hline                                                     
 \end{tabular}
 \label{tab:variability}
\end{table}

\section{Discussion} \label{sec:disc}
The origins of the X-ray emission in AGN hotspots has been discussed
extensively in the literature
\citep[e.g.,][]{2000TAVECCHIO,2001CELOTTI,hardcastle2004,GEOR2006,MINGO2017,
  MIGLIORI2020}. Initial studies of Pictor A in particular suggested
that a standard one-zone SSC model was incapable of explaining the
broadband SED, including an unusually high X-ray-to-radio flux ratio
\citep{WILSON2001}, as well as X-ray jet/counter-jet
  ratio and X-ray variability \citep{HARDCASTLE2016}.  Although
  already disfavored, we can test the IC-CMB hypothesis by comparing
  the extended X-ray spectrum with that in the radio, since if IC-CMB
  is responsible for the anomalous X-ray emission we expect to find
  that the X-ray spectral index is roughly equivalent to the
  low-frequency ($\sim$MHz) radio spectral index
  \citep{GEOR2006,JESTER2006}. The X-ray spectrum is not subject to
  the same requirements under a synchrotron origin, since the
  electrons producing the X-rays are distinct from those producing the
  radio emission.  In agreement with previous work
  \cite[e.g.][]{HARDCASTLE2016,THIMMAPPA2020} our results strongly
  disfavor an inverse Compton origin for the X-rays of the western
  Pictor A hotspot, based on the variability as well as the mismatch
  between the X-ray and radio spectral indices ($\Gamma_{X} =
  2.03\pm.03$ vs. $\Gamma_{R} = 1.74\pm.02$ respectively).

The minimum-energy (equipartition) magnetic field strength has been estimated at 0.36$\,mG$ \cite{ISOBE2017}, though later work has argued for a value 3-10 times larger under the hypothesis of highly magnetized substructures of scale less than a 100 pc \citep{ISOBE2020}. Under a synchrotron scenario the observed lower bound on the spectral cutoff energy of 41 keV corresponds to electron Lorentz factors $\gamma\sim1-3\times10^6$ using these values. The corresponding synchrotron cooling timescale $\tau_{syn}\sim (16/\gamma)(\mathrm{B}/\mathrm{1\,G})^{-2}$ \,yr using these values is 1-5 days. Since the cooling timescale is proportional to the square root of the corresponding photon energy, this 3-4 times shorter than that expected in the \emph{Chandra} band. In any case, these timescales are shorter than the minimum variability timescales observed with \emph{Chandra}, and correspond to very small spatial scales, possibly as small as light-days. In contrast the observed variability on month-year timescales implies $\sim$pcspatial scales of is in keeping with the observations of \cite{TINGAY2008}, who studied the Pictor A western hotspot using the Very Long Baseline Array (VLBA), finding multiple resolved pc-scale radio components within the hotspot. They suggested the X-rays could be produced by discrete substructures within the hotspot which are highly magnetized, which also could explain the observed excess emission in the IR \citep{ISOBE2017,ISOBE2020}.

The lack of high spatial resolution in the X-rays prevents us from spatially resolving the variability, but clearly the total flux we observe is integrated over many discrete regions, and therefore the low level (tens of percent) variability we see in aggregate translates to much larger-amplitude flaring within the actual small-scale region undergoing variability. Future coordinated monitoring campaigns with e.g., the VLBA and a next-generation sensitive and high-spatial-resolution X-ray telescope such as \emph{AXIS} may reveal these sites and the particle acceleration within them in action.

\section{Conclusions}

We have presented the results of a comprehensive analysis of the Western hotspot of Pictor A using a recent deep \emph{NuSTAR} observation combined with archival data from both \emph{NuSTAR} and \emph{Chandra}. We presented seperate (soft/hard) and combined fits to the wide X-ray band (0.5-20 keV) an danalyzed the X-ray emission for variability using the maximum likelihood method of \cite{MEYER2023}. The main results of these analyses are:

(i) We found that the \emph{NuSTAR} data alone is best fit by a simple power law of photon index $\Gamma=2.03\pm0.04$ and a normalization at 1 keV of $N_{1} = 118.9^{+8.7}_{-9.2}$. We do not find any evidence for a spectral break or cutoff energy in the hard X-ray band. The difference in normalization between the recent 2021 observations and previous epochs of both \emph{Chandra} and \emph{NuSTAR} data suggests variability within the western hotspot.

(ii) We found that the \emph{Chandra} data alone is best fit by a broken power law of photon indices $\Gamma_{1} = 1.95\pm0.02$ and $\Gamma_{2} = 2.15^{+0.10}_{-0.09}$, a break energy at $E_{brk} = 2.98^{+0.70}_{-0.51}$, and a normalization of $N_{1} = 95.0\pm0.7$ nJy at 1 keV.

(iii) We found that the combined \emph{Chandra} and \emph{NuSTAR} data are best fit by a broken power law model, with a best-fit photon indices of $\Gamma_{1} = 2.25^{+0.01}_{-0.03}$ and $\Gamma_{2} = 1.98^{+0.01}_{-0.01}$, a break energy of $E_{brk} = 0.76^{+0.07}_{-0.06} \mathrm{keV}$, and a normalization of $107.6^{+4.4}_{-3.8}$ nJy at 1 keV. However, the fit requires a significant offset factor of $f = 0.80$ for \emph{NuSTAR} relative to \emph{Chandra}. This suggests that the hotspot flux may have been higher during the recent \emph{NuSTAR} epoch and that interpretations of the joint fit must be made cautiously.

(iv) The maximum likelihood variability analysis of the \emph{Chandra} observations of the western hotspot finds statistically significant ($6.5\sigma$) variations in the 0.4-8 keV X-ray emission, confirming the results of \cite{HARDCASTLE2016} and \cite{THIMMAPPA2020}. We also confirm a significant ($3.3\sigma$) variation on a timescale of $\sim2$ months, placing an upper limit on the size of the varying region of $r<.01$ pc.

(v) The variability analysis of the \emph{NuSTAR} data alone is unable to find evidence of statistically significant variability in the hard, 3-20 keV X-ray flux, with a maximum significance of $1.74\sigma$. However, analyzing both the \emph{NuSTAR} and \emph{Chandra} data in the shared 3-8 keV energy range gives the most significant detection of variability yet, at $>8.2\sigma$.

Collectively, these results suggest a synchrotron origin for the hotspot X-ray emission, as the alternative beamed IC-CMB model is incapable of explaining either the rapid X-ray variability or the discrepancy between the measured X-ray spectral index ($\Gamma_{X} = 2.03\pm0.03$) and the expected low-frequency radio spectral index ($\Gamma_{R} = 1.74\pm0.02$).

The emerging picture of the hotspot is of multiple localized, highly magnetized, pc-scale substructures emitting X-ray predominantly via synchrotron radiation in agreement with previous research \citep[e.g.,][]{TINGAY2008,ISOBE2020}. This paradigm is capable of explaining the rapid X-ray variability and the difference in spectral index between the radio and X-rays. \textcolor{black}{Magnetic reconnection models are a possible explanation for \emph{in situ} acceleration in smaller regions and can produce electrons at the required TeV energies \citep[e.g.][]{WERNER2016,bodo2021}.} %However, this is only possible if the reconnection is triggered by ``hot" injection flows ($\gamma_{e}\sim100-1000$) as the magnetization alone cannot accelerate electrons beyond to the maximum energies.}
%we observed and in general magnetic reconnection models lack testable and falsifiable predictions.}

Follow-up observations of Pictor A with \emph{NuSTAR} might allow for a more significant detection of variability in the hard X-ray band and could confirm the flaring detected in the recent observations. The next major evolution will be the study of other nearby hotspots in the hard X-ray band. As mentioned previously, this is not a feasible task for \emph{NuSTAR} due to its angular resolution and the lower X-ray flux of most other hotspots relative to Pictor A's western hotspot. Next generation X-ray missions with higher angular resolution and sensitivity such as \emph{FORCE} \citep{NAKAZAWA2018}, \emph{HXMT} \citep{ZHANG2020}, and \emph{Athena} \citep{BARRET2023} should be capable of analyzing whether Pictor A's western hotspot is an outlier or a typical representative of the X-ray jet population.

\begin{acknowledgements}
This research has made use of data from the \emph{NuSTAR} mission led
by the California Institute of Technology and managed by NASA's Jet
Propulsion Laboratory, as well as software provided by the High Energy
Astrophysics Science Archive Research Center (HEASARC), a service of
the Astrophysics Science Division at NASA/GSFC and the High Energy
Astrophysics Division of the Smithsonian Astrophysical
Observatory. The observations were supported by the \emph{NuSTAR}
Guest Observer program. This research also made use of XSPEC
\citep{ARNAUD1996} and BXA \citep{BUCHNER2014}.
    
This paper employs a list of Chandra datasets, obtained by the Chandra
X-ray Observatory, contained in the Chandra Data Collection (CDC)
243~\dataset[doi:10.25574/cdc.243]{https://doi.org/10.25574/cdc.243}. This
research also made use of software provided by the Chandra X-ray
Center (CXC) in the application packages CIAO \citep{FRUSCIONE2006},
ChIPS \citep{GERMAIN2006}, and Sherpa
\citep{FREEMAN2001,DOE2007,BURKE2020}.

\end{acknowledgements}

\appendix
\setcounter{table}{0}
\renewcommand{\thetable}{A\arabic{table}}
\renewcommand{\thefigure}{A\arabic{figure}}
\setcounter{figure}{0}

\section{\emph{NuSTAR} background models} \label{app:nubgd}
%As described in Section \ref{sec:nubgd}, the \emph{NuSTAR} background model employed in this study is derived from \cite{WIK2014}.  
We briefly detail the various background components here; a more complete overview can be found in \citep{WIK2014}. 

\emph{\textbf{Internal background.}} This element models radiation resulting from the environment of \emph{NuSTAR}'s orbit and consists of two major components. The first is a featureless continuum, modeled as a broken power law with a break at 124 keV which dominates the background above 30 keV. The second is a collection of fluorescense and activation lines which dominates from approximately 20-30 keV, with weaker lines present up to $\sim100$ keV.

\emph{\textbf{Aperture background.}} The $\emph{NuSTAR}$ optical and focal plane modules are separated by an unenclosed mast. As a result, far away, off-axis stray light can leak into the aperture stop and strike the detector. Some of this stray light is blocked by the optics bench, resulting in a variable, spatially non-uniform background gradient across the detectors. The cosmic X-ray background (CXB) is the main contributor to this 'aperture' background, which dominates the background spectrum below 20 keV. \texttt{nuskybgd} uses the canonical \texttt{HEAO-1 A2} spectral model for the CXB \citep{GRUBER1999,REVN2003,CHURAZOV2007}.

\emph{\textbf{Scattered background.}} The \emph{NuSTAR} detectors are also affected by reflected and scattered X-rays from parts of the observatory. The main sources of these X-rays are the CXB, the Sun, and the Earth. These radiation sources are subdominant at all energies, but can contribute significantly at low energies. In particular, during periods of heightened solar activity, the solar background appears as a $\sim$1 keV thermal spectrum which can rival the aperture background radiation flux in the 3-5 keV energy range.

\emph{\textbf{fCXB.}} The `focused' CXB (fCXB) is a background radiation component resulting from unresolved sources within the FOV. The contribution of the fCXB is subdominant at all energies, but can contribute significantly below 15 keV, with a magnitude of approximately 10$\%$ of the CXB flux.

The \texttt{nuskybgd} package uses \texttt{XSPEC}\footnote{https://heasarc.gsfc.nasa.gov/xanadu/xspec} \citep{ARNAUD1996} to fit the spectrum of the entire background across the \emph{NuSTAR} detector modules, taking into account each component and any spatial variations present. \texttt{nuskybgd} groups the input spectrum to a minimum of 30 counts per bin and calculates the parameter normalizations using the $\chi^{2}$ statistic.  Within \texttt{XSPEC}, the background model is defined using a combination of four different models. \texttt{apbgd} and \texttt{fxapbgd} model the `aperture' and `focused aperture' or `fCXB' components and are both defined as cutoff power-law components (\texttt{cutoffpl}) with a photon index of $\Gamma = 1.29$ and a high-energy cutoff of $E = 41.13$ keV. The \texttt{line\_bgd} component models the fluorescense and activation line elements of the `internal' background component and the Solar thermal emission of the `scattered' background. It is comprised of 29 different Lorentzian (\texttt{lorentz}) line profiles and a thermal plasma component (\texttt{apec}) with temperature $kT = 1.15$. The final element is the \texttt{particle\_bgd} component which models the featureless continuum of the `internal' background as a broken power-law with photon indices of $\Gamma_{1} = -0.05, \Gamma_{2} = -0.85$ and a break enery of $E_{brk} = 124$ keV.

\begin{figure*}[!th]
    %\centeringl
    \gridline{
     \fig{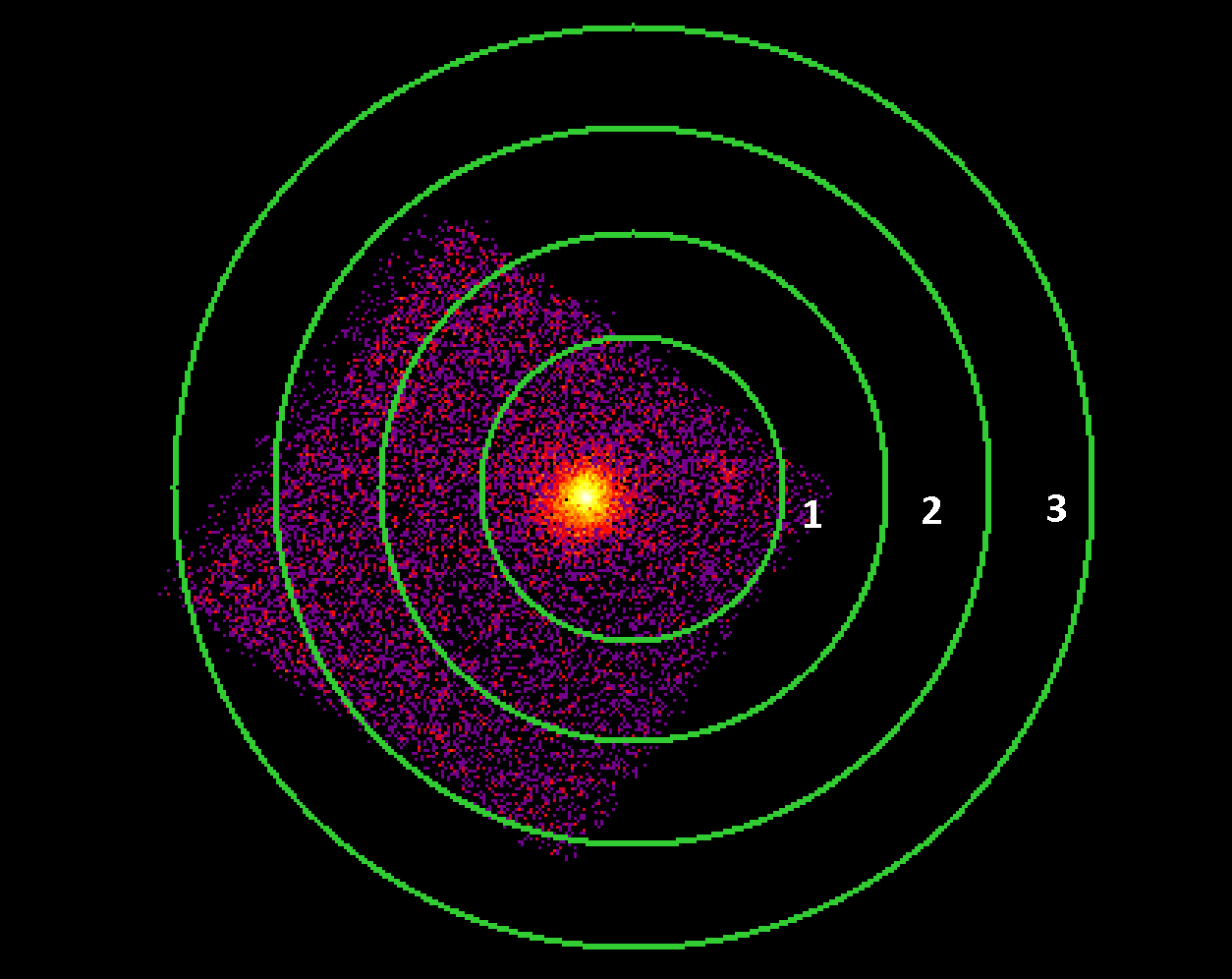}{0.60\textwidth}{}
    }
    \caption{An image of source Pictor A and its western hotspot generated from \emph{NuSTAR} epoch 1. The 3 green concentric circles define the annular background regions which are used by the \texttt{nuskybgd} package to map and calculate \emph{NuSTAR}'s asymmetric background radiation. The first annulus is centered around the source and hotspot to exclude its source emission. The two outer annuli are defined such that their inner radii are equal to the outer radii of the preceding annulus to maximize coverage of the image. The region sections which include pixels outside the detector's FOV do not impact the spectral analysis as the exposure value is 0.}
    \label{fig:nubg_ann}
\end{figure*}

We defined the background using 3 concentric annular regions which excluded emission from Pictor A and its jet and hotspot. An image example of these annular background regions for one of our \emph{NuSTAR} observations is shown in Figure \ref{fig:nubg_ann}. 
The spectral parameters of each model component such as the power-law photon indices and the line energies and widths are set by \texttt{nuskybgd} and fixed prior to fitting and do not change from observation to observation. Only the normalizations of these components are free to vary during the background fitting process. The normalizations of each component are recorded for each observation in Supplemental Table 1 in units of ph$\cdot\mathrm{cm}^{-2}\cdot\mathrm{s}^{-1} \cdot\mathrm{keV}^{-1}$.

\section{Results for individual \emph{Chandra} epochs}
In the main manuscript we have  reported the results of our analysis of the western hotspot spectrum using \emph{NuSTAR} data alone in Table \ref{tab:nustarlastresults} and the analysis of the total \emph{Chandra} and the combined \emph{NuSTAR}+\emph{Chandra} datsets in Table \ref{tab:lastresults}. Due to the number and length, the results for individual \emph{Chandra} epochs are reported here in Appendix Table \ref{tab:chandralastresults}.

\begin{sidewaystable}\centering
  \caption{Individual \emph{Chandra} epoch X-ray Spectral Fitting Results \label{tab:chandralastresults}}
\begin{tabular}{llllllBll|ccc}
  \hline
  \end{tabular}
\tablecomments{   * denotes the preferred model, according to the Bayesian evidence}
  \end{sidewaystable}


\begin{thebibliography}{}
\expandafter\ifx\csname natexlab\endcsname\relax\def\natexlab#1{#1}\fi
\providecommand{\url}[1]{\href{#1}{#1}}
\providecommand{\dodoi}[1]{doi:~\href{http://doi.org/#1}{\nolinkurl{#1}}}
\providecommand{\doeprint}[1]{\href{http://ascl.net/#1}{\nolinkurl{http://ascl.net/#1}}}
\providecommand{\doarXiv}[1]{\href{https://arxiv.org/abs/#1}{\nolinkurl{https://arxiv.org/abs/#1}}}


\bibitem[{{Akaike}(1974)}]{AKAIKE1974}
{Akaike}, H. 1974, IEEE Transactions on Automatic Control, 19, 716

\bibitem[{{Andrae} {et~al.}(2010){Andrae}, {Schulze-Hartung}, \& {Melchior}}]{ANDRAE2010}
{Andrae}, R., {Schulze-Hartung}, T., \& {Melchior}, P. 2010, arXiv e-prints, arXiv:1012.3754.
\newblock \doarXiv{1012.3754}

\bibitem[{{Arnaud}(1996)}]{ARNAUD1996}
{Arnaud}, K.~A. 1996, in Astronomical Society of the Pacific Conference Series, Vol. 101, Astronomical Data Analysis Software and Systems V, ed. G.~H. {Jacoby} \& J.~{Barnes}, 17

\bibitem[{{Arshakian} \& {Longair}(2000)}]{ARSHAKIAN2000}
{Arshakian}, T.~G., \& {Longair}, M.~S. 2000, \mnras, 311, 846, \dodoi{10.1046/j.1365-8711.2000.03098.x}

\bibitem[{{Barret} {et~al.}(2023){Barret}, {Albouys}, {Herder}, {Piro}, {Cappi}, {Huovelin}, {Kelley}, {Mas-Hesse}, {Paltani}, {Rauw}, {Rozanska}, {Svoboda}, {Wilms}, {Yamasaki}, {Audard}, {Bandler}, {Barbera}, {Barcons}, {Bozzo}, {Ceballos}, {Charles}, {Costantini}, {Dauser}, {Decourchelle}, {Duband}, {Duval}, {Fiore}, {Gatti}, {Goldwurm}, {Hartog}, {Jackson}, {Jonker}, {Kilbourne}, {Korpela}, {Macculi}, {Mendez}, {Mitsuda}, {Molendi}, {Pajot}, {Pointecouteau}, {Porter}, {Pratt}, {Pr{\^e}le}, {Ravera}, {Sato}, {Schaye}, {Shinozaki}, {Skup}, {Soucek}, {Thibert}, {Vink}, {Webb}, {Chaoul}, {Raulin}, {Simionescu}, {Torrejon}, {Acero}, {Branduardi-Raymont}, {Ettori}, {Finoguenov}, {Grosso}, {Kaastra}, {Mazzotta}, {Miller}, {Miniutti}, {Nicastro}, {Sciortino}, {Yamaguchi}, {Beaumont}, {Cucchetti}, {D'Andrea}, {Eckart}, {Ferrando}, {Kammoun}, {Lotti}, {Mesnager}, {Natalucci}, {Peille}, {de Plaa}, {Ardellier}, {Argan}, {Bellouard}, {Carron}, {Cavazzuti}, {Fiorini}, {Khosropanah}, {Martin}, {Perry}, {Pinsard},
  {Pradines}, {Rigano}, {Roelfsema}, {Schwander}, {Torrioli}, {Ullom}, {Vera}, {Villegas}, {Zuchniak}, {Brachet}, {Cicero}, {Doriese}, {Durkin}, {Fioretti}, {Geoffray}, {Jacques}, {Kirsch}, {Smith}, {Adams}, {Gloaguen}, {Hoogeveen}, {van der Hulst}, {Kiviranta}, {van der Kuur}, {Ledot}, {van Leeuwen}, {van Loon}, {Lyautey}, {Parot}, {Sakai}, {van Weers}, {Abdoelkariem}, {Adam}, {Adami}, {Aicardi}, {Akamatsu}, {Alonso}, {Amato}, {Andr{\'e}}, {Angelinelli}, {Anon-Cancela}, {Anvar}, {Atienza}, {Attard}, {Auricchio}, {Balado}, {Bancel}, {Barusso}, {Bascu{\~n}an}, {Bernard}, {Berrocal}, {Blin}, {Bonino}, {Bonnet}, {Bonny}, {Boorman}, {Boreux}, {Bounab}, {Boutelier}, {Boyce}, {Brienza}, {Bruijn}, {Bulgarelli}, {Calarco}, {Callanan}, {Campello}, {Camus}, {Canourgues}, {Capobianco}, {Cardiel}, {Castellani}, {Cheatom}, {Chervenak}, {Chiarello}, {Clerc}, {Clerc}, {Cobo}, {Coeur-Joly}, {Coleiro}, {Colonges}, {Corcione}, {Coriat}, {Coynel}, {Cuttaia}, {D'Ai}, {D'anca}, {Dadina}, {Daniel}, {Dauner}, {DeNigris},
  {Dercksen}, {DiPirro}, {Doumayrou}, {Dubbeldam}, {Dupieux}, {Dupourqu{\'e}}, {Durand}, {Eckert}, {Eiriz}, {Ercolani}, {Etcheverry}, {Finkbeiner}, {Fiocchi}, {Fossecave}, {Franssen}, {Frericks}, {Gabici}, {Gant}, {Gao}, {Gastaldello}, {Genolet}, {Ghizzardi}, {Gil}, {Giovannini}, {Godet}, {Gomez-Elvira}, {Gonzalez}, {Gonzalez}, {Gottardi}, {Granat}, {Gros}, {Guignard}, {Hieltjes}, {Hurtado}, {Irwin}, {Jacquey}, {Janiuk}, {Jaubert}, {Jim{\'e}nez}, {Jolly}, {Jourdan}, {Julien}, {Kedziora}, {Korb}, {Kreykenbohm}, {K{\"o}nig}, {Langer}, {Laudet}, {Laurent}, {Laurenza}, {Lesrel}, {Ligori}, {Lorenz}, {Luminari}, {Maffei}, {Maisonnave}, {Marelli}, {Massonet}, {Maussang}, {Melchor}, {Le Mer}, {Millan}, {Millerioux}, {Mineo}, {Minervini}, {Molin}, {Monestes}, {Montinaro}, {Mot}, {Murat}, {Nagayoshi}, {Naz{\'e}}, {Nogu{\`e}s}, {Pailot}, {Panessa}, {Parodi}, {Petit}, {Piconcelli}, {Pinto}, {Plaza}, {Plaza}, {Poyatos}, {Prouv{\'e}}, {Ptak}, {Puccetti}, {Puccio}, {Ramon}, {Reina}, {Rioland}, {Rodriguez}, {Roig}, {Rollet},
  {Roncarelli}, {Roudil}, {Rudnicki}, {Sanisidro}, {Sciortino}, {Silva}, {Sordet}, {Soto-Aguilar}, {Spizzi}, {Surace}, {Fern{\'a}ndez S{\'a}nchez}, {Taralli}, {Terrasa}, {Terrier}, {Todaro}, {Ubertini}, {Uslenghi}, {de Vaate}, {Vaccaro}, {Varisco}, {Varni{\`e}re}, {Vibert}, {Vidriales}, {Villa}, {Vodopivec}, {Volpe}, {de Vries}, {Wakeham}, {Walmsley}, {Wise}, {de Wit}, \& {Wo{\'z}niak}}]{BARRET2023}
{Barret}, D., {Albouys}, V., {Herder}, J.-W.~d., {et~al.} 2023, Experimental Astronomy, 55, 373, \dodoi{10.1007/s10686-022-09880-7}

\bibitem[{{Blandford} \& {K{\"o}nigl}(1979)}]{BLANDFORD1979}
{Blandford}, R.~D., \& {K{\"o}nigl}, A. 1979, \apj, 232, 34, \dodoi{10.1086/157262}

\bibitem[{{Bodo} {et~al.}(2021){Bodo}, {Tavecchio}, \& {Sironi}}]{bodo2021}
{Bodo}, G., {Tavecchio}, F., \& {Sironi}, L. 2021, \mnras, 501, 2836, \dodoi{10.1093/mnras/staa3620}

\bibitem[{{Breiding} {et~al.}(2017){Breiding}, {Meyer}, {Georganopoulos}, {Keenan}, {DeNigris}, \& {Hewitt}}]{BREI2017}
{Breiding}, P., {Meyer}, E.~T., {Georganopoulos}, M., {et~al.} 2017, \apj, 849, 95, \dodoi{10.3847/1538-4357/aa907a}

\bibitem[{{Breiding} {et~al.}(2023){Breiding}, {Meyer}, {Georganopoulos}, {Reddy}, {Kollmann}, \& {Roychowdhury}}]{BREIDING2023}
---. 2023, \mnras, 518, 3222, \dodoi{10.1093/mnras/stac3081}

\bibitem[{{Buchner}(2016)}]{BUCHNER2016}
{Buchner}, J. 2016, {BXA: Bayesian X-ray Analysis}, Astrophysics Source Code Library, record ascl:1610.011.
\newblock \doeprint{1610.011}

\bibitem[{{Buchner}(2021)}]{BUCHNER2021}
---. 2021, The Journal of Open Source Software, 6, 3001, \dodoi{10.21105/joss.03001}

\bibitem[{{Buchner} {et~al.}(2014){Buchner}, {Georgakakis}, {Nandra}, {Hsu}, {Rangel}, {Brightman}, {Merloni}, {Salvato}, {Donley}, \& {Kocevski}}]{BUCHNER2014}
{Buchner}, J., {Georgakakis}, A., {Nandra}, K., {et~al.} 2014, \aap, 564, A125, \dodoi{10.1051/0004-6361/201322971}

\bibitem[{{Burke} {et~al.}(2020){Burke}, {Laurino}, {Wmclaugh}, {Dtnguyen2}, {Marie-Terrell}, {G{\"u}nther}, {Budynkiewicz}, {Siemiginowska}, {Aldcroft}, {Deil}, {Sip{\H{o}}cz}, {Leinweber}, \& {Todd}}]{BURKE2020}
{Burke}, D., {Laurino}, O., {Wmclaugh}, {et~al.} 2020, {sherpa/sherpa: Sherpa 4.12.1}, 4.12.1,  Zenodo, \dodoi{10.5281/zenodo.3944985}

\bibitem[{{Cash}(1979)}]{CASH1979}
{Cash}, W. 1979, \apj, 228, 939, \dodoi{10.1086/156922}

\bibitem[{{Celotti} {et~al.}(2001){Celotti}, {Ghisellini}, \& {Chiaberge}}]{2001CELOTTI}
{Celotti}, A., {Ghisellini}, G., \& {Chiaberge}, M. 2001, \mnras, 321, L1, \dodoi{10.1046/j.1365-8711.2001.04160.x}

\bibitem[{{Chartas} {et~al.}(2000){Chartas}, {Worrall}, {Birkinshaw}, {Cresitello-Dittmar}, {Cui}, {Ghosh}, {Harris}, {Hooper}, {Jauncey}, {Kim}, {Lovell}, {Mathur}, {Schwartz}, {Tingay}, {Virani}, \& {Wilkes}}]{CHARTAS2000}
{Chartas}, G., {Worrall}, D.~M., {Birkinshaw}, M., {et~al.} 2000, \apj, 542, 655, \dodoi{10.1086/317049}

\bibitem[{{Churazov} {et~al.}(2007){Churazov}, {Sunyaev}, {Revnivtsev}, {Sazonov}, {Molkov}, {Grebenev}, {Winkler}, {Parmar}, {Bazzano}, {Falanga}, {Gros}, {Lebrun}, {Natalucci}, {Ubertini}, {Roques}, {Bouchet}, {Jourdain}, {Kn{\"o}dlseder}, {Diehl}, {Budtz-Jorgensen}, {Brandt}, {Lund}, {Westergaard}, {Neronov}, {T{\"u}rler}, {Chernyakova}, {Walter}, {Produit}, {Mowlavi}, {Mas-Hesse}, {Domingo}, {Gehrels}, {Kuulkers}, {Kretschmar}, \& {Schmidt}}]{CHURAZOV2007}
{Churazov}, E., {Sunyaev}, R., {Revnivtsev}, M., {et~al.} 2007, \aap, 467, 529, \dodoi{10.1051/0004-6361:20066230}

\bibitem[{{Dabhade} {et~al.}(2020){Dabhade}, {Mahato}, {Bagchi}, {Saikia}, {Combes}, {Sankhyayan}, {R{\"o}ttgering}, {Ho}, {Gaikwad}, {Raychaudhury}, {Vaidya}, \& {Guiderdoni}}]{dab2020}
{Dabhade}, P., {Mahato}, M., {Bagchi}, J., {et~al.} 2020, \aap, 642, A153, \dodoi{10.1051/0004-6361/202038344}

\bibitem[{{Doe} {et~al.}(2007){Doe}, {Nguyen}, {Stawarz}, {Refsdal}, {Siemiginowska}, {Burke}, {Evans}, {Evans}, {McDowell}, {Houck}, \& {Nowak}}]{DOE2007}
{Doe}, S., {Nguyen}, D., {Stawarz}, C., {et~al.} 2007, in Astronomical Society of the Pacific Conference Series, Vol. 376, Astronomical Data Analysis Software and Systems XVI, ed. R.~A. {Shaw}, F.~{Hill}, \& D.~J. {Bell}, 543

\bibitem[{{Erlund} {et~al.}(2007){Erlund}, {Fabian}, {Blundell}, {Moss}, \& {Ballantyne}}]{ERLUND2007}
{Erlund}, M.~C., {Fabian}, A.~C., {Blundell}, K.~M., {Moss}, C., \& {Ballantyne}, D.~R. 2007, \mnras, 379, 498, \dodoi{10.1111/j.1365-2966.2007.11962.x}

\bibitem[{Esch {et~al.}(2004)Esch, Connors, Karovska, \& van Dyk}]{esch2004image}
Esch, D.~N., Connors, A., Karovska, M., \& van Dyk, D.~A. 2004, The Astrophysical Journal, 610, 1213

\bibitem[{{Fabian}(2012)}]{FABI2012}
{Fabian}, A.~C. 2012, \araa, 50, 455, \dodoi{10.1146/annurev-astro-081811-125521}

\bibitem[{{Fanaroff} \& {Riley}(1974)}]{FANA1974}
{Fanaroff}, B.~L., \& {Riley}, J.~M. 1974, \mnras, 167, 31P, \dodoi{10.1093/mnras/167.1.31P}

\bibitem[{{Freeman} {et~al.}(2001){Freeman}, {Doe}, \& {Siemiginowska}}]{FREEMAN2001}
{Freeman}, P., {Doe}, S., \& {Siemiginowska}, A. 2001, in Society of Photo-Optical Instrumentation Engineers (SPIE) Conference Series, Vol. 4477, Astronomical Data Analysis, ed. J.-L. {Starck} \& F.~D. {Murtagh}, 76--87, \dodoi{10.1117/12.447161}

\bibitem[{{Fruscione} {et~al.}(2006){Fruscione}, {McDowell}, {Allen}, {Brickhouse}, {Burke}, {Davis}, {Durham}, {Elvis}, {Galle}, {Harris}, {Huenemoerder}, {Houck}, {Ishibashi}, {Karovska}, {Nicastro}, {Noble}, {Nowak}, {Primini}, {Siemiginowska}, {Smith}, \& {Wise}}]{FRUSCIONE2006}
{Fruscione}, A., {McDowell}, J.~C., {Allen}, G.~E., {et~al.} 2006, in Society of Photo-Optical Instrumentation Engineers (SPIE) Conference Series, Vol. 6270, Society of Photo-Optical Instrumentation Engineers (SPIE) Conference Series, ed. D.~R. {Silva} \& R.~E. {Doxsey}, 62701V, \dodoi{10.1117/12.671760}

\bibitem[{{Garmire} {et~al.}(2003){Garmire}, {Bautz}, {Ford}, {Nousek}, \& {Ricker}}]{GARMIRE2003}
{Garmire}, G.~P., {Bautz}, M.~W., {Ford}, P.~G., {Nousek}, J.~A., \& {Ricker}, George~R., J. 2003, in Society of Photo-Optical Instrumentation Engineers (SPIE) Conference Series, Vol. 4851, X-Ray and Gamma-Ray Telescopes and Instruments for Astronomy., ed. J.~E. {Truemper} \& H.~D. {Tananbaum}, 28--44, \dodoi{10.1117/12.461599}

\bibitem[{{Georganopoulos} \& {Kazanas}(2003)}]{GEOR2003}
{Georganopoulos}, M., \& {Kazanas}, D. 2003, \apjl, 589, L5, \dodoi{10.1086/375796}

\bibitem[{{Georganopoulos} {et~al.}(2006){Georganopoulos}, {Perlman}, {Kazanas}, \& {McEnery}}]{GEOR2006}
{Georganopoulos}, M., {Perlman}, E.~S., {Kazanas}, D., \& {McEnery}, J. 2006, \apjl, 653, L5, \dodoi{10.1086/510452}

\bibitem[{{Germain} {et~al.}(2006){Germain}, {Milaszewski}, {McLaughlin}, {Miller}, {Evans}, {Evans}, \& {Burke}}]{GERMAIN2006}
{Germain}, G., {Milaszewski}, R., {McLaughlin}, W., {et~al.} 2006, in Astronomical Society of the Pacific Conference Series, Vol. 351, Astronomical Data Analysis Software and Systems XV, ed. C.~{Gabriel}, C.~{Arviset}, D.~{Ponz}, \& S.~{Enrique}, 57

\bibitem[{{Gordon} \& {Arnaud}(2021)}]{ARNAUD2021}
{Gordon}, C., \& {Arnaud}, K. 2021, {PyXspec: Python interface to XSPEC spectral-fitting program}, Astrophysics Source Code Library, record ascl:2101.014.
\newblock \doeprint{2101.014}

\bibitem[{{Gruber} {et~al.}(1999){Gruber}, {Matteson}, {Peterson}, \& {Jung}}]{GRUBER1999}
{Gruber}, D.~E., {Matteson}, J.~L., {Peterson}, L.~E., \& {Jung}, G.~V. 1999, \apj, 520, 124, \dodoi{10.1086/307450}

\bibitem[{{Hardcastle} {et~al.}(2002){Hardcastle}, {Birkinshaw}, {Cameron}, {Harris}, {Looney}, \& {Worrall}}]{hardcastle2002}
{Hardcastle}, M.~J., {Birkinshaw}, M., {Cameron}, R.~A., {et~al.} 2002, \apj, 581, 948, \dodoi{10.1086/344409}

\bibitem[{{Hardcastle} \& {Croston}(2005)}]{HARDCASTLE2005}
{Hardcastle}, M.~J., \& {Croston}, J.~H. 2005, \mnras, 363, 649, \dodoi{10.1111/j.1365-2966.2005.09469.x}

\bibitem[{{Hardcastle} {et~al.}(2004){Hardcastle}, {Harris}, {Worrall}, \& {Birkinshaw}}]{hardcastle2004}
{Hardcastle}, M.~J., {Harris}, D.~E., {Worrall}, D.~M., \& {Birkinshaw}, M. 2004, \apj, 612, 729, \dodoi{10.1086/422808}

\bibitem[{{Hardcastle} {et~al.}(2016){Hardcastle}, {Lenc}, {Birkinshaw}, {Croston}, {Goodger}, {Marshall}, {Perlman}, {Siemiginowska}, {Stawarz}, \& {Worrall}}]{HARDCASTLE2016}
{Hardcastle}, M.~J., {Lenc}, E., {Birkinshaw}, M., {et~al.} 2016, \mnras, 455, 3526, \dodoi{10.1093/mnras/stv2553}

\bibitem[{{Harris} {et~al.}(2003){Harris}, {Biretta}, {Junor}, {Perlman}, {Sparks}, \& {Wilson}}]{HARRIS2003}
{Harris}, D.~E., {Biretta}, J.~A., {Junor}, W., {et~al.} 2003, \apjl, 586, L41, \dodoi{10.1086/374773}

\bibitem[{{Harris} {et~al.}(1994){Harris}, {Carilli}, \& {Perley}}]{harris1994}
{Harris}, D.~E., {Carilli}, C.~L., \& {Perley}, R.~A. 1994, \nat, 367, 713, \dodoi{10.1038/367713a0}

\bibitem[{{Harris} \& {Krawczynski}(2006)}]{HARRIS2006}
{Harris}, D.~E., \& {Krawczynski}, H. 2006, \araa, 44, 463, \dodoi{10.1146/annurev.astro.44.051905.092446}

\bibitem[{{Harris} \& {Krawczynski}(2007)}]{HARRIS2007}
{Harris}, D.~E., \& {Krawczynski}, H. 2007, in Revista Mexicana de Astronomia y Astrofisica Conference Series, Vol.~27, Revista Mexicana de Astronomia y Astrofisica, vol. 27, 188, \dodoi{10.48550/arXiv.astro-ph/0604527}

\bibitem[{{Harris} {et~al.}(2010){Harris}, {Massaro}, \& {Cheung}}]{HARRIS2010}
{Harris}, D.~E., {Massaro}, F., \& {Cheung}, C.~C. 2010, in American Institute of Physics Conference Series, Vol. 1248, X-ray Astronomy 2009; Present Status, Multi-Wavelength Approach and Future Perspectives, ed. A.~{Comastri}, L.~{Angelini}, \& M.~{Cappi}, 355--358, \dodoi{10.1063/1.3475257}

\bibitem[{{Harrison} {et~al.}(2013){Harrison}, {Craig}, {Christensen}, {Hailey}, {Zhang}, {Boggs}, {Stern}, {Cook}, {Forster}, {Giommi}, {Grefenstette}, {Kim}, {Kitaguchi}, {Koglin}, {Madsen}, {Mao}, {Miyasaka}, {Mori}, {Perri}, {Pivovaroff}, {Puccetti}, {Rana}, {Westergaard}, {Willis}, {Zoglauer}, {An}, {Bachetti}, {Barri{\`e}re}, {Bellm}, {Bhalerao}, {Brejnholt}, {Fuerst}, {Liebe}, {Markwardt}, {Nynka}, {Vogel}, {Walton}, {Wik}, {Alexander}, {Cominsky}, {Hornschemeier}, {Hornstrup}, {Kaspi}, {Madejski}, {Matt}, {Molendi}, {Smith}, {Tomsick}, {Ajello}, {Ballantyne}, {Balokovi{\'c}}, {Barret}, {Bauer}, {Blandford}, {Brandt}, {Brenneman}, {Chiang}, {Chakrabarty}, {Chenevez}, {Comastri}, {Dufour}, {Elvis}, {Fabian}, {Farrah}, {Fryer}, {Gotthelf}, {Grindlay}, {Helfand}, {Krivonos}, {Meier}, {Miller}, {Natalucci}, {Ogle}, {Ofek}, {Ptak}, {Reynolds}, {Rigby}, {Tagliaferri}, {Thorsett}, {Treister}, \& {Urry}}]{HARRISON2013}
{Harrison}, F.~A., {Craig}, W.~W., {Christensen}, F.~E., {et~al.} 2013, \apj, 770, 103, \dodoi{10.1088/0004-637X/770/2/103}

\bibitem[{Harrison {et~al.}(2013)Harrison, Craig, Christensen, Hailey, Zhang, Boggs, Stern, Cook, Forster, Giommi, Grefenstette, Kim, Kitaguchi, Koglin, Madsen, Mao, Miyasaka, Mori, Perri, Pivovaroff, Puccetti, Rana, Westergaard, Willis, Zoglauer, An, Bachetti, Barri{\`{e}}re, Bellm, Bhalerao, Brejnholt, Fuerst, Liebe, Markwardt, Nynka, Vogel, Walton, Wik, Alexander, Cominsky, Hornschemeier, Hornstrup, Kaspi, Madejski, Matt, Molendi, Smith, Tomsick, Ajello, Ballantyne, Balokovi{\'{c}}, Barret, Bauer, Blandford, Brandt, Brenneman, Chiang, Chakrabarty, Chenevez, Comastri, Dufour, Elvis, Fabian, Farrah, Fryer, Gotthelf, Grindlay, Helfand, Krivonos, Meier, Miller, Natalucci, Ogle, Ofek, Ptak, Reynolds, Rigby, Tagliaferri, Thorsett, Treister, \& Urry}]{Harrison_2013}
Harrison, F.~A., Craig, W.~W., Christensen, F.~E., {et~al.} 2013, The Astrophysical Journal, 770, 103, \dodoi{10.1088/0004-637x/770/2/103}

\bibitem[{{HI4PI Collaboration} {et~al.}(2016){HI4PI Collaboration}, {Ben Bekhti}, {Fl{\"o}er}, {Keller}, {Kerp}, {Lenz}, {Winkel}, {Bailin}, {Calabretta}, {Dedes}, {Ford}, {Gibson}, {Haud}, {Janowiecki}, {Kalberla}, {Lockman}, {McClure-Griffiths}, {Murphy}, {Nakanishi}, {Pisano}, \& {Staveley-Smith}}]{H4PI2016}
{HI4PI Collaboration}, {Ben Bekhti}, N., {Fl{\"o}er}, L., {et~al.} 2016, \aap, 594, A116, \dodoi{10.1051/0004-6361/201629178}

\bibitem[{{Isobe} {et~al.}(2017){Isobe}, {Koyama}, {Kino}, {Wada}, {Nakagawa}, {Matsuhara}, {Niinuma}, \& {Tashiro}}]{ISOBE2017}
{Isobe}, N., {Koyama}, S., {Kino}, M., {et~al.} 2017, \apj, 850, 193, \dodoi{10.3847/1538-4357/aa94c9}

\bibitem[{{Isobe} {et~al.}(2020){Isobe}, {Sunada}, {Kino}, {Koyama}, {Tashiro}, {Nagai}, \& {Pearson}}]{ISOBE2020}
{Isobe}, N., {Sunada}, Y., {Kino}, M., {et~al.} 2020, \apj, 899, 17, \dodoi{10.3847/1538-4357/ab9d1c}

\bibitem[{{Jester} {et~al.}(2006){Jester}, {Harris}, {Marshall}, \& {Meisenheimer}}]{JESTER2006}
{Jester}, S., {Harris}, D.~E., {Marshall}, H.~L., \& {Meisenheimer}, K. 2006, \apj, 648, 900, \dodoi{10.1086/505962}

\bibitem[{Kashyap {et~al.}(2017)Kashyap, van Dyk, McKeough, Primini, Jerius, Gowrishankar, Siemiginowska, \& Zezas}]{Kashyap2017XrayingTE}
Kashyap, V.~L., van Dyk, D.~A., McKeough, K., {et~al.} 2017, Proceedings of the International Astronomical Union, 12, 284

\bibitem[{{Knuth} {et~al.}(2014){Knuth}, {Habeck}, {Malakar}, {Mubeen}, \& {Placek}}]{KNUTH2014}
{Knuth}, K.~H., {Habeck}, M., {Malakar}, N.~K., {Mubeen}, A.~M., \& {Placek}, B. 2014, arXiv e-prints, arXiv:1411.3013.
\newblock \doarXiv{1411.3013}

\bibitem[{{Kraft} {et~al.}(2002){Kraft}, {Forman}, {Jones}, {Murray}, {Hardcastle}, \& {Worrall}}]{KRAFT2002}
{Kraft}, R.~P., {Forman}, W.~R., {Jones}, C., {et~al.} 2002, \apj, 569, 54, \dodoi{10.1086/339062}

\bibitem[{{Madsen} {et~al.}(2017){Madsen}, {Beardmore}, {Forster}, {Guainazzi}, {Marshall}, {Miller}, {Page}, \& {Stuhlinger}}]{MADSEN2017}
{Madsen}, K.~K., {Beardmore}, A.~P., {Forster}, K., {et~al.} 2017, \aj, 153, 2, \dodoi{10.3847/1538-3881/153/1/2}

\bibitem[{{Madsen} {et~al.}(2014){Madsen}, {Harrison}, {An}, {Boggs}, {Christensen}, {Cook}, {Craig}, {Forster}, {Fuerst}, {Grefenstette}, {Hailey}, {Kitaguchi}, {Markwardt}, {Mao}, {Miyasaka}, {Rana}, {Stern}, {Zhang}, {Zoglauer}, {Walton}, \& {Westergaard}}]{MADSEN2014}
{Madsen}, K.~K., {Harrison}, F.~A., {An}, H., {et~al.} 2014, in Society of Photo-Optical Instrumentation Engineers (SPIE) Conference Series, Vol. 9144, Space Telescopes and Instrumentation 2014: Ultraviolet to Gamma Ray, ed. T.~{Takahashi}, J.-W.~A. {den Herder}, \& M.~{Bautz}, 91441P, \dodoi{10.1117/12.2056643}

\bibitem[{{Madsen} {et~al.}(2015){Madsen}, {Harrison}, {Markwardt}, {An}, {Grefenstette}, {Bachetti}, {Miyasaka}, {Kitaguchi}, {Bhalerao}, {Boggs}, {Christensen}, {Craig}, {Forster}, {Fuerst}, {Hailey}, {Perri}, {Puccetti}, {Rana}, {Stern}, {Walton}, {J{\o}rgen Westergaard}, \& {Zhang}}]{MADSEN2015}
{Madsen}, K.~K., {Harrison}, F.~A., {Markwardt}, C.~B., {et~al.} 2015, \apjs, 220, 8, \dodoi{10.1088/0067-0049/220/1/8}

\bibitem[{{Marshall} {et~al.}(2010){Marshall}, {Hardcastle}, {Birkinshaw}, {Croston}, {Evans}, {Landt}, {Lenc}, {Massaro}, {Perlman}, {Schwartz}, {Siemiginowska}, {Stawarz}, {Urry}, \& {Worrall}}]{MARSHALL2010}
{Marshall}, H.~L., {Hardcastle}, M.~J., {Birkinshaw}, M., {et~al.} 2010, \apjl, 714, L213, \dodoi{10.1088/2041-8205/714/2/L213}

\bibitem[{{McNamara} {et~al.}(2009){McNamara}, {Kazemzadeh}, {Rafferty}, {B{\^\i}rzan}, {Nulsen}, {Kirkpatrick}, \& {Wise}}]{MCNAMARA2009A}
{McNamara}, B.~R., {Kazemzadeh}, F., {Rafferty}, D.~A., {et~al.} 2009, \apj, 698, 594, \dodoi{10.1088/0004-637X/698/1/594}

\bibitem[{{Meisenheimer} {et~al.}(1997){Meisenheimer}, {Yates}, \& {Roeser}}]{MEISENHEIMER1997}
{Meisenheimer}, K., {Yates}, M.~G., \& {Roeser}, H.~J. 1997, \aap, 325, 57

\bibitem[{{Meyer} {et~al.}(2016){Meyer}, {Sparks}, {Georganopoulos}, {Anderson}, {van der Marel}, {Biretta}, {Sohn}, {Chiaberge}, {Perlman}, \& {Norman}}]{Meyer2016}
{Meyer}, E.~T., {Sparks}, W.~B., {Georganopoulos}, M., {et~al.} 2016, \apj, 818, 195, \dodoi{10.3847/0004-637X/818/2/195}

\bibitem[{{Meyer} {et~al.}(2023){Meyer}, {Shaik}, {Tang}, {Reid}, {Reddy}, {Breiding}, {Georganopoulos}, {Chiaberge}, {Perlman}, {Clautice}, {Sparks}, {DeNigris}, \& {Trevor}}]{MEYER2023}
{Meyer}, E.~T., {Shaik}, A., {Tang}, Y., {et~al.} 2023, Nature Astronomy, \dodoi{10.1038/s41550-023-01983-1}

\bibitem[{{Migliori} {et~al.}(2007){Migliori}, {Grandi}, {Palumbo}, {Brunetti}, \& {Stanghellini}}]{MIGLIORI2007}
{Migliori}, G., {Grandi}, P., {Palumbo}, G. G.~C., {Brunetti}, G., \& {Stanghellini}, C. 2007, \apj, 668, 203, \dodoi{10.1086/520870}

\bibitem[{{Migliori} {et~al.}(2020){Migliori}, {Orienti}, {Coccato}, {Brunetti}, {D'Ammando}, {Mack}, \& {Prieto}}]{MIGLIORI2020}
{Migliori}, G., {Orienti}, M., {Coccato}, L., {et~al.} 2020, \mnras, 495, 1593, \dodoi{10.1093/mnras/staa1214}

\bibitem[{{Mingo} {et~al.}(2017){Mingo}, {Hardcastle}, {Ineson}, {Mahatma}, {Croston}, {Dicken}, {Evans}, {Morganti}, \& {Tadhunter}}]{MINGO2017}
{Mingo}, B., {Hardcastle}, M.~J., {Ineson}, J., {et~al.} 2017, \mnras, 470, 2762, \dodoi{10.1093/mnras/stx1307}

\bibitem[{{Mullin} {et~al.}(2008){Mullin}, {Riley}, \& {Hardcastle}}]{MULLIN2008}
{Mullin}, L.~M., {Riley}, J.~M., \& {Hardcastle}, M.~J. 2008, \mnras, 390, 595, \dodoi{10.1111/j.1365-2966.2008.13534.x}

\bibitem[{{Nakazawa} {et~al.}(2018){Nakazawa}, {Mori}, {Tsuru}, {Ueda}, {Awaki}, {Fukazawa}, {Ishida}, {Matsumoto}, {Murakami}, {Okajima}, {Takahashi}, {Tsunemi}, \& {Zhang}}]{NAKAZAWA2018}
{Nakazawa}, K., {Mori}, K., {Tsuru}, T.~G., {et~al.} 2018, in Society of Photo-Optical Instrumentation Engineers (SPIE) Conference Series, Vol. 10699, Space Telescopes and Instrumentation 2018: Ultraviolet to Gamma Ray, ed. J.-W.~A. {den Herder}, S.~{Nikzad}, \& K.~{Nakazawa}, 106992D, \dodoi{10.1117/12.2309344}

\bibitem[{{Reddy} {et~al.}(2021){Reddy}, {Georganopoulos}, \& {Meyer}}]{2021ApJS..253...37R}
{Reddy}, K., {Georganopoulos}, M., \& {Meyer}, E.~T. 2021, \apjs, 253, 37, \dodoi{10.3847/1538-4365/abd8d7}

\bibitem[{Reddy {et~al.}(2023)Reddy, Georganopoulos, Meyer, Keenan, \& Kollmann}]{Reddy_2023}
Reddy, K., Georganopoulos, M., Meyer, E.~T., Keenan, M., \& Kollmann, K.~E. 2023, The Astrophysical Journal Supplement Series, 265, 8, \dodoi{10.3847/1538-4365/aca321}

\bibitem[{{Revnivtsev} {et~al.}(2003){Revnivtsev}, {Gilfanov}, {Sunyaev}, {Jahoda}, \& {Markwardt}}]{REVN2003}
{Revnivtsev}, M., {Gilfanov}, M., {Sunyaev}, R., {Jahoda}, K., \& {Markwardt}, C. 2003, \aap, 411, 329, \dodoi{10.1051/0004-6361:20031386}

\bibitem[{{Sambruna} {et~al.}(2004){Sambruna}, {Gambill}, {Maraschi}, {Tavecchio}, {Cerutti}, {Cheung}, {Urry}, \& {Chartas}}]{SAMBRUNA2004}
{Sambruna}, R.~M., {Gambill}, J.~K., {Maraschi}, L., {et~al.} 2004, \apj, 608, 698, \dodoi{10.1086/383124}

\bibitem[{{Sambruna} {et~al.}(2001){Sambruna}, {Maraschi}, {Tavecchio}, {Cheung}, {Urry}, {Chartas}, {Scarpa}, \& {Pesce}}]{SAMBRUNA2001}
{Sambruna}, R.~M., {Maraschi}, L., {Tavecchio}, F., {et~al.} 2001, in American Institute of Physics Conference Series, Vol. 587, Gamma 2001: Gamma-Ray Astrophysics, ed. S.~{Ritz}, N.~{Gehrels}, \& C.~R. {Shrader}, 256--260, \dodoi{10.1063/1.1419409}

\bibitem[{Schwarz(1978)}]{SCHWARZ1978}
Schwarz, G. 1978, The Annals of Statistics, 6, 461 , \dodoi{10.1214/aos/1176344136}

\bibitem[{{Smith} {et~al.}(2001){Smith}, {Brickhouse}, {Liedahl}, \& {Raymond}}]{SMITH2001}
{Smith}, R.~K., {Brickhouse}, N.~S., {Liedahl}, D.~A., \& {Raymond}, J.~C. 2001, \apjl, 556, L91, \dodoi{10.1086/322992}

\bibitem[{{Snios} {et~al.}(2019){Snios}, {Wykes}, {Nulsen}, {Kraft}, {Meyer}, {Birkinshaw}, {Worrall}, {Hardcastle}, {Roediger}, {Forman}, \& {Jones}}]{SNIOS2019}
{Snios}, B., {Wykes}, S., {Nulsen}, P. E.~J., {et~al.} 2019, \apj, 871, 248, \dodoi{10.3847/1538-4357/aafaf3}

\bibitem[{Stein {et~al.}(2015)Stein, van Dyk, Kashyap, \& Siemiginowska}]{stein2015detecting}
Stein, N.~M., van Dyk, D.~A., Kashyap, V.~L., \& Siemiginowska, A. 2015, The Astrophysical Journal, 813, 66

\bibitem[{{Sun} {et~al.}(2018){Sun}, {Yang}, {Rieger}, {Liu}, \& {Aharonian}}]{SUN2018}
{Sun}, X.-N., {Yang}, R.-Z., {Rieger}, F.~M., {Liu}, R.-Y., \& {Aharonian}, F. 2018, \aap, 612, A106, \dodoi{10.1051/0004-6361/201731716}

\bibitem[{{Sunada} {et~al.}(2022){Sunada}, {Morimoto}, {Tashiro}, {Terada}, {Katsuda}, {Sato}, {Tateishi}, \& {Sasaki}}]{SUNADA2022}
{Sunada}, Y., {Morimoto}, A., {Tashiro}, M.~S., {et~al.} 2022, \pasj, 74, 602, \dodoi{10.1093/pasj/psac022}

\bibitem[{{Tavecchio} {et~al.}(2000){Tavecchio}, {Maraschi}, {Sambruna}, \& {Urry}}]{2000TAVECCHIO}
{Tavecchio}, F., {Maraschi}, L., {Sambruna}, R.~M., \& {Urry}, C.~M. 2000, \apjl, 544, L23, \dodoi{10.1086/317292}

\bibitem[{{Thimmappa} {et~al.}(2020){Thimmappa}, {Stawarz}, {Marchenko}, {Balasubramaniam}, {Cheung}, \& {Siemiginowska}}]{THIMMAPPA2020}
{Thimmappa}, R., {Stawarz}, {\L}., {Marchenko}, V., {et~al.} 2020, \apj, 903, 109, \dodoi{10.3847/1538-4357/abb605}

\bibitem[{{Tingay} {et~al.}(2008){Tingay}, {Lenc}, {Brunetti}, \& {Bondi}}]{TINGAY2008}
{Tingay}, S.~J., {Lenc}, E., {Brunetti}, G., \& {Bondi}, M. 2008, \aj, 136, 2473, \dodoi{10.1088/0004-6256/136/6/2473}

\bibitem[{{Weisskopf} {et~al.}(2000){Weisskopf}, {Tananbaum}, {Van Speybroeck}, \& {O'Dell}}]{WEISSKOPF2000A}
{Weisskopf}, M.~C., {Tananbaum}, H.~D., {Van Speybroeck}, L.~P., \& {O'Dell}, S.~L. 2000, in Society of Photo-Optical Instrumentation Engineers (SPIE) Conference Series, Vol. 4012, X-Ray Optics, Instruments, and Missions III, ed. J.~E. {Truemper} \& B.~{Aschenbach}, 2--16, \dodoi{10.1117/12.391545}

\bibitem[{{Werner} {et~al.}(2016){Werner}, {Uzdensky}, {Cerutti}, {Nalewajko}, \& {Begelman}}]{WERNER2016}
{Werner}, G.~R., {Uzdensky}, D.~A., {Cerutti}, B., {Nalewajko}, K., \& {Begelman}, M.~C. 2016, \apjl, 816, L8, \dodoi{10.3847/2041-8205/816/1/L8}

\bibitem[{{Werner} {et~al.}(2012){Werner}, {Murphy}, {Livingston}, {Gorjian}, {Jones}, {Meier}, \& {Lawrence}}]{WERNER2012}
{Werner}, M.~W., {Murphy}, D.~W., {Livingston}, J.~H., {et~al.} 2012, \apj, 759, 86, \dodoi{10.1088/0004-637X/759/2/86}

\bibitem[{{Wik} {et~al.}(2014){Wik}, {Hornstrup}, {Molendi}, {Madejski}, {Harrison}, {Zoglauer}, {Grefenstette}, {Gastaldello}, {Madsen}, {Westergaard}, {Ferreira}, {Kitaguchi}, {Pedersen}, {Boggs}, {Christensen}, {Craig}, {Hailey}, {Stern}, \& {Zhang}}]{WIK2014}
{Wik}, D.~R., {Hornstrup}, A., {Molendi}, S., {et~al.} 2014, \apj, 792, 48, \dodoi{10.1088/0004-637X/792/1/48}

\bibitem[{{Wilson} {et~al.}(2001){Wilson}, {Young}, \& {Shopbell}}]{WILSON2001}
{Wilson}, A.~S., {Young}, A.~J., \& {Shopbell}, P.~L. 2001, \apj, 547, 740, \dodoi{10.1086/318412}

\bibitem[{{Worrall}(2009)}]{WORRALL2009}
{Worrall}, D.~M. 2009, \aapr, 17, 1, \dodoi{10.1007/s00159-008-0016-7}

\bibitem[{{Worrall} {et~al.}(2012){Worrall}, {Birkinshaw}, {Young}, {Momtahan}, {Fosbury}, {Morganti}, {Tadhunter}, \& {Verdoes Kleijn}}]{WORRALL2012}
{Worrall}, D.~M., {Birkinshaw}, M., {Young}, A.~J., {et~al.} 2012, \mnras, 424, 1346, \dodoi{10.1111/j.1365-2966.2012.21320.x}

\bibitem[{{Zhang} {et~al.}(2020){Zhang}, {Li}, {Lu}, {Song}, {Xu}, {Liu}, {Chen}, {Cao}, {Bu}, {Chang}, {Chen}, {Chen}, {Chen}, {Chen}, {Chen}, {Cui}, {Cui}, {Deng}, {Dong}, {Du}, {Fu}, {Gao}, {Gao}, {Gao}, {Ge}, {Gu}, {Guan}, {Gungor}, {Guo}, {Han}, {Hu}, {Huang}, {Huo}, {Jia}, {Jiang}, {Jiang}, {Jin}, {Jin}, {Li}, {Li}, {Li}, {Li}, {Li}, {Li}, {Li}, {Li}, {Li}, {Li}, {Li}, {Liang}, {Liao}, {Liu}, {Liu}, {Liu}, {Liu}, {Liu}, {Liu}, {Lu}, {Lu}, {Luo}, {Ma}, {Meng}, {Nang}, {Nie}, {Ou}, {Qu}, {Sai}, {Shang}, {Shen}, {Sun}, {Tan}, {Tao}, {Tuo}, {Wang}, {Wang}, {Wang}, {Wang}, {Wang}, {Wang}, {Wang}, {Wen}, {Wu}, {Wu}, {Wu}, {Xiao}, {Xiong}, {Yan}, {Yang}, {Yang}, {Yang}, {Yi}, {Yuan}, {Zhang}, {Zhang}, {Zhang}, {Zhang}, {Zhang}, {Zhang}, {Zhang}, {Zhang}, {Zhang}, {Zhang}, {Zhang}, {Zhang}, {Zhang}, {Zhang}, {Zhang}, {Zhang}, {Zhang}, {Zhang}, {Zhang}, {Zhang}, {Zhao}, {Zhao}, {Zheng}, {Zhou}, {Zhu}, {Zhu}, {Zhuang}, \& {Insight-HXMT Team}}]{ZHANG2020}
{Zhang}, S.-N., {Li}, T., {Lu}, F., {et~al.} 2020, Science China Physics, Mechanics, and Astronomy, 63, 249502, \dodoi{10.1007/s11433-019-1432-6}



\end{thebibliography}
\end{document}